\documentclass[manuscript,screen]{acmart}

\usepackage{tabularx}
\usepackage{csvsimple}
\AtBeginDocument{%
  \providecommand\BibTeX{{%
    \normalfont B\kern-0.5em{\scshape i\kern-0.25em b}\kern-0.8em\TeX}}}

\setcopyright{acmcopyright}
\copyrightyear{2018}
\acmYear{2018}
\acmDOI{10.1145/1122445.1122456}

\acmBooktitle{Woodstock '18: ACM Symposium on Neural Gaze Detection,
  June 03--05, 2018, Woodstock, NY}
\acmPrice{15.00}
\acmISBN{978-1-4503-XXXX-X/18/06}

\begin{document}

\title{Leveraging the Defects Life Cycle to Label Affected Versions and Defective Classes}

\author{Bailey Vandehei}
\email{bvandehe@calpoly.edu }
\affiliation{%
  \institution{California Polytechnic State University}
  \city{San Luis Obispo}
  \state{California}
}

\author{Daniel Alencar da Costa}
\email{danielcalencar@otago.ac.nz}
\affiliation{%
  \institution{University of Otago}
  \city{Dunedin}
  \country{New Zealand}}

\author{Davide Falessi}
\email{falessi@ing.uniroma2.it}
\affiliation{%
  \institution{University of Rome Tor Vergata}
  \city{Rome}
  \country{Italy}
}

\renewcommand{\shortauthors}{Vandehei et al.}

\begin{abstract}
Two recent studies explicitly recommend labeling defective classes in releases using the affected versions (AV) available in issue trackers (e.g., JIRA). This practice is coined as the {\em realistic approach}. However, no study has investigated whether it is feasible to rely on AVs. For example, how available and consistent is the AV information on existing issue trackers? Additionally, no study has attempted to retrieve AVs when they are unavailable. The aim our study is threefold: 1) to measure the proportion of defects for which the realistic method is usable, 2) to propose a method for retrieving the AVs of a defect, thus making  the realistic approach usable when AVs are unavailable, 3) to compare the accuracy of the proposed method versus three SZZ implementations.  
The assumption of our proposed method is that defects have a stable life cycle in terms of the proportion of the number of versions affected by the defects before discovering and fixing these defects. Results related to 212 open-source projects from the Apache ecosystem, featuring a total of about 125,000 defects, reveal that the realistic method cannot be used in the majority (51\%) of defects. Therefore, it is important to develop automated methods to retrieve AVs. Results related to 76 open-source projects from the Apache ecosystem, featuring a total of about 6,250,000 classes, affected by 60,000 defects, and spread over 4,000 versions and 760,000 commits, reveal that the proportion of the number of versions between defect discovery and fix is pretty stable (STDV \textless 2)---across the defects of the same project. Moreover, the proposed method resulted significantly more accurate than all three SZZ implementations in (i) retrieving AVs, (ii) labeling classes as defective, and (iii) in developing defects repositories to perform feature selection. Thus, when the realistic method is unusable, the proposed method is a valid automated alternative to SZZ for retrieving the origin of a defect. Finally, given the low accuracy of SZZ, researchers should consider re-executing the studies that have used SZZ as an oracle and, in general, should prefer selecting projects with a high proportion of available and consistent AVs. 
\end{abstract}


\maketitle
\section{Introduction}\label{sec:intro}
The manner in which defects are introduced into code, and
the sheer volume of defects in software, are typically beyond the capability and resources of most development teams~\cite{DBLP:conf/icse/Tantithamthavorn16,DBLP:journals/corr/abs-1801-10270,TSE:Nam:2017,DBLP:journals/tse/Kamei/2012,MSR:Ghotra:2017,kondo2019impact}.
Due to this problem, researchers have explored machine learning approaches to predict (1) whether a defect is likely
to occur in a software module (i.e, a binary response); and (2) how many post-deployment defects are likely to occur (a count response)~\cite{TSE:Nam:2017,DBLP:journals/infsof/FuMS16,DBLP:journals/tse/Kim/2008,DBLP:conf/ase/Nam/2015,OOPSLA:Madhavan:2007,ASE:Shivaji:2009,ASE:Jiang:2013,chen2019predicting,fu2017revisiting,scanniello2013class}. 

Defect prediction models are built using complexity features (i.e., lines of code, Halstead metrics, McCabe’s complexity, and CK metrics)~\cite{TSE:Basili:1996,halstead1977elements,mccabe1976complexity} and process features~\cite{ICSE:rahman:2013}. Predicting the occurrence of defects is useful because a development team could better focus the limited testing effort.

Before any defect prediction can be performed, it is important to create a repository containing the features and the associated defects. Our work focuses on the automated methods for the creation of defect prediction datasets. We are interested in methods for establishing the origin of a defect.
Researchers provided means to create \cite{DBLP:journals/spe/GaoKWS11, Zimmermann:2007:PDE:1268984.1269057,menzies2012promise}, collect  \cite{DBLP:conf/sigsoft/FalessiM18} and select \cite{DBLP:journals/ese/GousiosS14, DBLP:conf/msr/RozenbergBKPBPM16, Nagappan:2013:DSE:2491411.2491415} datasets for associating software defects to process and product features. However, existing research has shown that the general quality of software defect datasets are not without flaws \cite{Kim:2011:DND:1985793.1985859, Herzig:2013:IBI:2486788.2486840, Rahman:2013:SSV:2491411.2491418, Tantithamthavorn:2015:IMP:2818754.2818852, Bird:2009:FBB:1595696.1595716}. For example, \citet{Bird:2009:FBB:1595696.1595716} demonstrated the existence of a non-negligible bias in the features that are used to build defect prediction models.  \citet{Tantithamthavorn:2015:IMP:2818754.2818852} have also shown that cleaning the datasets prior to performing defect predictions can increase the ability to better identify the defective modules. Indeed, the general accuracy of a defect prediction model depends on the quality of the underlying datasets \cite{Kochhar:2014:PBB:2642937.2642997, Shepperd:2013}.

One main limitation of defect prediction models is the granularity of the predictions (e.g., whether a module is defective or not), which is often too coarse to be useful~\cite{kamei2016defect}. To face this limitation, researchers have explored {\em Just-In-Time} (JIT) defect predictions~\cite{fan2019impact}, in which the goal of the prediction is to indicate whether a newly produced commit will be defective or clean. Nevertheless, JIT prediction models can only be feasible if the exact origins of a defect are known~\cite{fan2019impact}. 

To identify the origins of a defect, researchers have proposed the SZZ approach \cite{Sliwerski:2005:CIF:1082983.1083147}. However, the state-of-art of the SZZ approach is far from being ideal~\cite{CostaMSKCH17,DBLP:journals/infsof/Rodriguez-Perez18,Rodriguez-Perez:2018:BDO:3239235.3267436}. For example, \citet{CostaMSKCH17} highlighted that current SZZ implementations cannot determine the origins of defects that were fixed by solely adding code. Additionally, SZZ is also incapable of identifying the origins of 
defects of the regression type~\cite{vaswani2010approach}. Finally, \citet{rodriguezbugs} revealed that only a significant minority of defects can have their origins traceable in the source code repository, thus limiting the applicability of SZZ. 

Two recent studies~\cite{CostaMSKCH17,Yatish2019} suggest the use of {\em affected versions} (AVs) available in defect reports---which can be provided in issue trackers such as JIRA---to better label defective modules, instead of solely relying on SZZ. However, these studies also hint that the availability of AVs is scarce~\cite{CostaMSKCH17,Yatish2019}, i.e., only a few defect reports provide AVs. In this work, we propose a first-of-its-kind method to retrieve AV. The method, if used in combination with fix commit information, is used also to label defective classes.  
To achieve our goal, we first investigate the extent to which AVs are usable, i.e., available and consistent, in open-source projects. Second, we propose, evaluate, and compare novel and automated methods for retrieving AVs, including the earliest possible AV (i.e., the origin of the defect).  
Our intuition is that {\em defects have a stable life cycle} in terms of the proportion of the number of versions required to discover and to fix a defect. The idea is that defects that quickly manifest themselves as a fault (i.e., report creation) are easiest to find and fix than defects that are dormant over several releases~\cite{chen_nagappan_shihab_hassan_2014,AalokSnoring2019}. This is because developers need to identify the change that induced the defect to fix the defect. Our assumption is that: the older the defect-inducing change is, the longer it takes for it to be identified and fixed. The assumption of the stability of defects' life-cycle seem to have analogies with {\em diseases'} life-cycle \cite{rodenhuis2010dengue}.

Our results obtained in 212 Apache open-source projects reveal that AV information is lacking in the majority of defects (51\%). Therefore, it is important to investigate automated methods for retrieving AVs. Our results obtained in 76 Apache open-source projects demonstrate that our proposed method is more accurate than previously proposed SZZ methods in terms of retrieving AVs. Additionally, our methods are more accurate in labeling classes as defective and in developing defects datasets for performing feature selection.

The remainder of this paper is structured as follows. We explain the background material and related work in Section 2. In Section 3, we describe our study design. We present our obtained results in Section 4. In Section 5, we discuss our results. We explain the threats to validity of this study in Section 6, while we provide our conclusions in Section 7.

\section{Related Work \& Background}\label{sec:related}
\label{sec:RelatedWork}

We provide the key concepts to understand our research context in this section. 

\begin{figure}
    \centering
    \includegraphics[width=\textwidth,keepaspectratio]
    {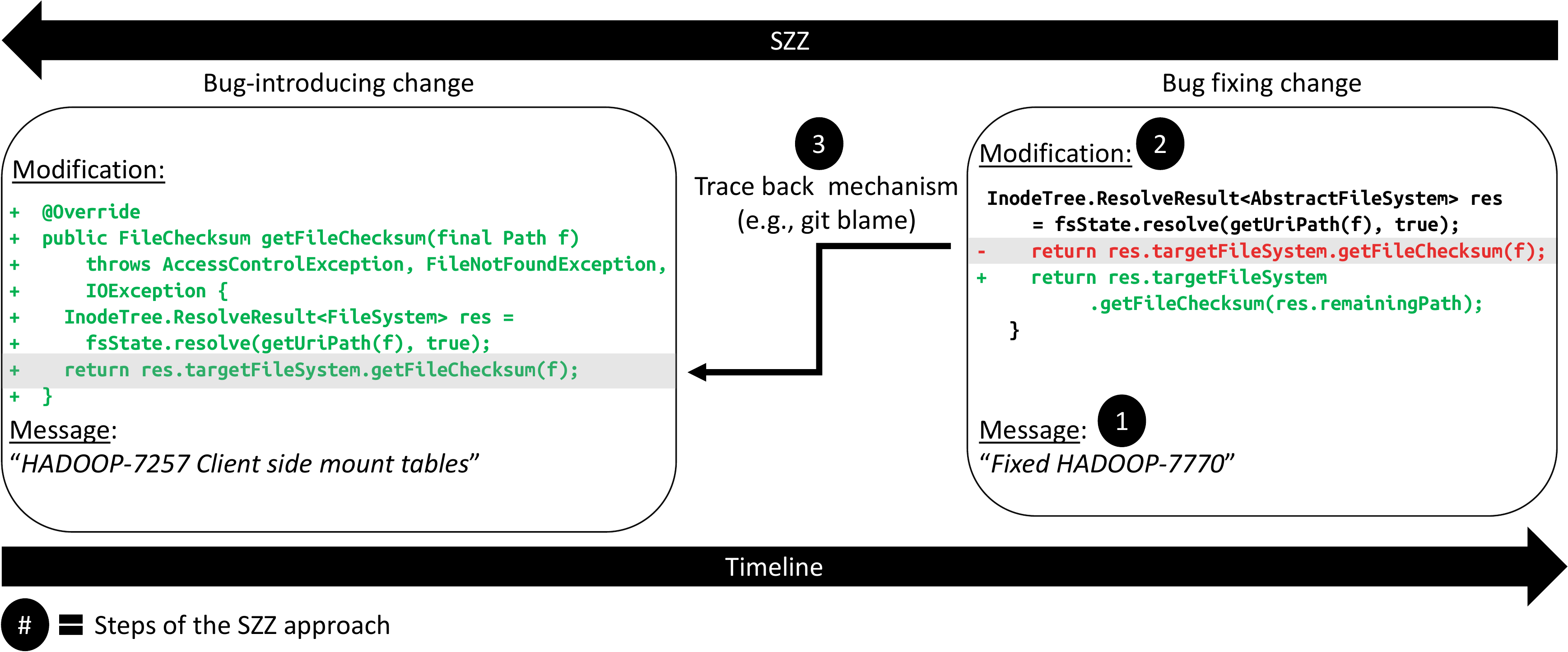}
    \caption{An example of the SZZ approach. Step 1 consists on identifying the defect-fixing changes. Step 2 localizes the faulty code, which is the code removed in the defect-fixing change. Finally, in Step 3, SZZ traces the code history to find the defect-introducing changes.}
    \label{fig:szz}
\end{figure}

\citet{Sliwerski:2005:CIF:1082983.1083147} proposed the first implementation of the SZZ approach, which strove to find the origins of a defect (i.e., the defect-introducing changes). SZZ exploits the versioning system annotation mechanism (e.g. git blame) to determine---for the source code lines that have been changed in a defect fix---when they have last been changed before the fix. The SZZ approach consists of three main steps. We demonstrate these steps by using the  HADOOP-7770\footnote{\url{https://issues.apache.org/jira/browse/HADOOP-7770}} defect as an example (shown in Figure~\ref{fig:szz}). HADOOP-7770 was caused because the developers used the wrong object to provide a file path, which incurred a \texttt{FileNotFoundException}. 
{\bfseries Step 1} of SZZ (shown in Figure~\ref{fig:szz}) consists of finding the change that fixed the defect (i.e., the defect-fixing change). In the case of HADOOP-7770, the defect-fixing change was performed in change \texttt{1190532}\footnote{\url{http://svn.apache.org/viewvc?view=revision&revision=1190541}} by changing \texttt{getFileChecksum(f)} to \texttt{getFileChecksum(res.remainingPath)}.
SZZ can use several mechanisms to find defect-fixing changes~\cite{Sliwerski:2005:CIF:1082983.1083147}.\footnote{A popular approach to identify defect-fixing changes is to use simple heuristics, such as searching for the ``fix'' or ``fixed'' keywords in a change log~\cite{eyolfson2011time}. However, the SZZ implementations used in this work search for defect IDs within change logs for identifying the defect-fixing changes.} 
Afterwards, in {\bfseries Step~2}, SZZ analyzes the diff patch of the defect-fixing change to locate the {\em faulty code}. In this step, SZZ assumes that the code removed in a patch is the code that expresses the defect. 
In the case of HADOOP-7770, the removed code in the diff patch was the \texttt{getFileChecksum(f);} code. Finally, once the {\em faulty code} has been identified, SZZ traces the code history to find when the {\em faulty code} was introduced (i.e., {\bfseries Step~3}). {\bfseries Step~3} of SZZ can be implemented by using, for example, the \texttt{blame} operation that is present in most Version Control Systems (VCSs, such as Git or Subversion). In Figure~\ref{fig:szz}, SZZ uses the {\em git blame} command to find the change, \texttt{1100026}\footnote{\url{http://svn.apache.org/viewvc?view=revision&revision=1100026}}, which is the change that introduced the \texttt{getFileChecksum(f);} and, hence, the code that potentially introduced the defect (i.e., the defect-introducing change).

Several other studies strove to estimate the origin of defects. \citet{Kim:2006:AIB:1169218.1169308} presented algorithms to automatically and accurately identify defect-introducing changes which improved over SZZ. \citet{CostaMSKCH17} proposed three criteria and evaluated five SZZ implementations. They concluded that current SZZ implementations still lack mechanisms to accurately identify defect-introducing changes. \citet{Yatish2019} presented the realistic approach (i.e., using AVs) to estimate the origin of a defect. This approach relies on the use of the AV and is the main motivation of the present work.  \citet{DBLP:conf/wcre/NetoCK18} found that 19.9\% of lines that are removed during a fix are related to refactoring operations and, therefore, their respective defect-introducing changes are likely false positives. \citet{DBLP:conf/sigsoft/FalessiM18} presented the Pilot Defects Prediction Dataset Maker (PDPDM), a desktop application to measure metrics for use in defect prediction. PDPDM avoids the use of outdated datasets, and it allows researchers and practitioners to create defect datasets without  writing any code. \citet{Rodriguez-Perez:2018:BDO:3239235.3267436} investigated the complex phenomenon of defect introduction and defect fix. They showed that less than 30\% of defects can actually be traced to its origins by assuming that ``a given defect was introduced by the lines of code that were modified to fix it".
Our research complements the prior research in defect introduction by providing methods to retrieve the AVs. AVs can then be used to evaluate or improve approaches such as SZZ~\cite{CostaMSKCH17}. 

Extensive research has been invested in building and evaluating datasets for defect prediction.~\citet{DBLP:journals/tse/Shepperd/2013} investigated five studies that have used the NASA dataset for building defect prediction models. The goal of their work was to verify whether the different versions of the NASA dataset yield consistent results. \citet{DBLP:journals/tse/Shepperd/2013} observed that different versions of the same dataset (e.g., NASA) may produce different results for defect prediction and, therefore, researchers should be cautious before selecting a dataset. \citet{DBLP:conf/ase/Nam/2015} proposed the CLA and CLAMI approaches to automatically label unlabelled defect prediction datasets, relieving researchers from the manual effort. The approaches work based on the magnitude of metric values and obtain average prediction performances of around 0.64 (F-measure) and 0.72 (AUC). 

Other studies focused on how to select repositories to mine. \citet{Nagappan:2013:DSE:2491411.2491415} combined ideas from representativeness and diversity, and introduced a measure called sample coverage, which is the percentage of projects in a population that are similar to the given sample. They concluded that studies should discuss the target population of the research (universe) and dimensions that potentially can influence the outcomes of the research (space). \citet{DBLP:journals/ese/GousiosS14} proposed the Alitheia Core analysis platform, which pre-processes repository data into an intermediate format that allows researchers to provide custom analysis tools. \citet{DBLP:conf/msr/RozenbergBKPBPM16} proposed RepoGrams to support researchers in qualitatively comparing and contrasting software projects over time using a set of software metrics. RepoGrams uses an extensible, metrics-based, visualization model that can be adapted to a variety of analyses. \citet{Falessi:2017:SSF:3200492.3200515} presented STRESS, a semi-automated and fully replicable approach that allows researchers to select projects by configuring the desired level of diversity, fit, and quality.

A significant effort has been invested in measuring the noise in defects repositories and its impact on the follow-up analyses. \citet{Bird:2009:FBB:1595696.1595716} found that  bias is a critical problem that threatens both the effectiveness of processes that rely on biased datasets to build prediction models and the generalizability of hypotheses tested on biased data. \citet{Kim:2011:DND:1985793.1985859} measured the impact of noise on defect prediction models and provided guidelines for acceptable noise levels. They also proposed a noise detection and elimination algorithm to address this problem. However, the noise studied and removed is supposed to be random. \citet{Herzig:2013:IBI:2486788.2486840} reported that 39\% of files marked as defective actually never had a defect. They discussed the impact of this misclassification on earlier studies and recommended manual data validation for future studies. \citet{Rahman:2013:SSV:2491411.2491418} showed that size always matters just as much as bias direction, and in fact, much more than bias direction when considering information-retrieval measures such as AUCROC and F-score. This indicates that, at least for prediction models, even when dealing with sampling bias, simply finding larger samples can sometimes be sufficient. \citet{Tantithamthavorn:2015:IMP:2818754.2818852}  found that: (1) issue report mislabelling is not random; (2) precision is rarely impacted by mislabelled issue reports, suggesting that practitioners can rely on the accuracy of modules labelled as defective by models that are trained using noisy data; (3) however, models trained on noisy data typically achieve about 60\% of the recall of models trained on clean data. Complementary to the aforementioned studies, we measure the extent of noise (i.e. classes mislabeled) and its impact on analyzing a repository in terms of features selection. 

Another line of research in defect prediction has proposed the usage of a machine learning model to predict whether an upcoming change (i.e., commit) is defective or clean~\cite{DBLP:journals/tse/Kim/2008,DBLP:journals/tse/Kamei/2012,DBLP:conf/qrs/Yang/2015,DBLP:journals/emse/Kamei/2016,DBLP:conf/fse/Yang/2016,DBLP:conf/msr/Fukushima/2014}. This area of research was eventually coined as {\em Just-in-time} defect prediction (JIT). \citet{DBLP:journals/tse/Kim/2008} proposed the usage of JIT models in their seminal work. In order to label their datasets, the authors used the output from SZZ. \citet{DBLP:conf/msr/Fukushima/2014} and \citet{DBLP:journals/emse/Kamei/2016} advanced the area and explored the usage of cross-project JIT models to help software projects without enough historical data to build their own models. In our work, we aim at retrieving AVs and verifying whether our methods can improve the accuracy on labeling defective classes. It is worth to note that we do not use our approaches for building JIT models as we envision to do so in future work. 

Other researchers suggest that prediction models should be used, and validated according to their ability, to ranking modules with respect to their defect proneness~\cite{Yang:IDEAL:2012,Yang:TRel:2014,Yu:TRel:2019,Yu:SANER:2019,Panichella::2016}. Yang et al.~\cite{Yang:IDEAL:2012} proposed the use of Learning-to-Rank (LTR) algorithms to rank the defect proneness of software modules in their seminal work. Later, Yang et al.~\cite{Yang:TRel:2014} expanded the their seminal work to (i) apply the LTR method to datasets containing multiple releases; (ii) to perform more comparisons between the LTR method and other algorithms (e.g., algorithms based on least squares); and (iii) to investigate additional metrics to build and evaluate defect prediction models. Panichella et al.~\cite{Panichella:GECCO:2016} proposed the usage of genetic alrogithms to tune the parameters of defect prediction to maximize the ratio between the number of defects and the required effort to inspect these defects. This ratio between defects found and required effort to inspect them can  be coined as `cost.' The authors observed that their approach significantly outperformed traditional models. Yu et al.~\cite{Yu:TRel:2019} proposed a cost-sensitive support vector machine algorithm to improve ranking-oriented defect prediction models. For example, predicting the wrong rank for a module that contains a higher number of defects is worse than predicting the wrong rank for a module with less number of defects. Therefore, Yu et al.'s~\cite{Yu:TRel:2019} approach is sensitive to such costs. The authors found that their approach outperformed approaches that are not cost-sensitive. Our work can potentially complement the rank-oriented defect prediction models in the sense that retrieving AVs can improve the labeling of defective modules. Therefore, a potential future work is to build rank-oriented defect prediction models using datasets labelled by our proposed approach.

\section{Research Questions}\label{sec:design}

\newcommand{\rqone}{Are AVs available and consistent?}
\newcommand{\rqtwo}{Do methods have different accuracy for labeling affected  versions?}
\newcommand{\rqthree}{Do methods have different accuracy for labeling defective classes?}
\newcommand{\rqfour}{Do methods lead to selecting different features?}

In this paper we investigate four research questions:
\begin{itemize}
\item \textbf{RQ1: \rqone} Two recent SZZ studies \cite{CostaMSKCH17,Yatish2019} recommend using affected versions (AV) to identify the origin of a defect and, hence, create defect datasets. However, how often do developers actually provide AVs in defect reports?  In this research question, we investigate the extent to which AVs are usable, i.e., whether they are available and consistent. 
\item \textbf{RQ2: \rqtwo} If AV of a defect is not available then it must be retrieved. In this research question, we compare the accuracy of ten different methods for labeling versions as affected or not by a defect. The ten methods consist of three baseline methods belonging to the SZZ family and seven new methods which leverage the lifecycle information of defects.
\item \textbf{RQ3: \rqthree} In order to mine a dataset of defects, it is important to have information about which class in which version is defective. Therefore, in this research question, we investigate which methods have the highest accuracy for labeling defective classes. We use commits' information to trace classes to defect-fixing changes and hence labeling specific classes as defective or not. We then merged this information with the information about affected versions (RQ2) to label classes  in specific versions as defective or not.  In other words, in this research question we observe the accuracy of the realistic approach \cite{Yatish2019}, in labeling classes, when the AVs are retrieved by different methods. This investigation is important as the accuracy of mining activities is highly correlated with the correctness of the datasets that are used \cite{Kochhar:2014:PBB:2642937.2642997, Shepperd:2013}.
\item \textbf{RQ4: \rqfour}  To measure the level of impact on practitioners of the use of different methods, in this research question, we investigate the accuracy of methods in leading to accurate feature selection. Specifically, feature selection is the activity of identifying what features contribute the most in a prediction model for predicting whether a class is defective \cite{Hall1998, kondo2019impact,chen2019predicting,fu2017revisiting,scanniello2013class}. If a feature is selected as important, then it is strongly associated with the defectiveness of classes. As such, these important features can provide practitioners and researchers with knowledge on how to avoid future defects \cite{Zimmermann:2007:PDE:1268984.1269057, Hall1998,DBLP:journals/sigsoft/SliwerskiZZ05, DBLP:conf/msr/EyolfsonTL11, DBLP:conf/msr/AsaduzzamanBRS12, DBLP:conf/icse/RahmanD11, DBLP:conf/msr/WeissPZZ07, DBLP:conf/csmr/BernardiCLPD12,DBLP:conf/msr/RahmanBD10,DBLP:conf/icse/KimZWZ07}. However, inaccurate datasets may lead to the identification of the wrong features and hence to ineffective suggestions on how to avoid defects. 
\end{itemize}
As the feature selection accuracy depends on the accuracy of labeling defective classes, which in turn depends on the accuracy of labeling affected versions, then RQ2 results impact RQ3 results which, in turn, impacts RQ4 results. Since the actual extent of the differences across methods performances in specific research questions is currently unknown and could vary due to minor factors, then the existence of correlations across research questions does not decrease the value of each research question. For instance, the set of features identified (RQ4) could not vary, despite the variation of the methods' accuracy in defective class labeling (RQ3), due to the use of the specific set of features. For example, one feature could be selected regardless of the method used to develop the dataset, since the other features are uncorrelated to it anyhow.

Before reporting on the design and results of each of our research questions, we present the concepts that are shared across research questions.

We define a failure as an incorrect or unexpected result provided by the system or, more generally speaking, when the system behaves in an unintended way. Note that a failure is always caused by a defect, but a defect causes a failure only under specific conditions.

\begin{figure}
  \includegraphics[width=0.9\textwidth,keepaspectratio]
	{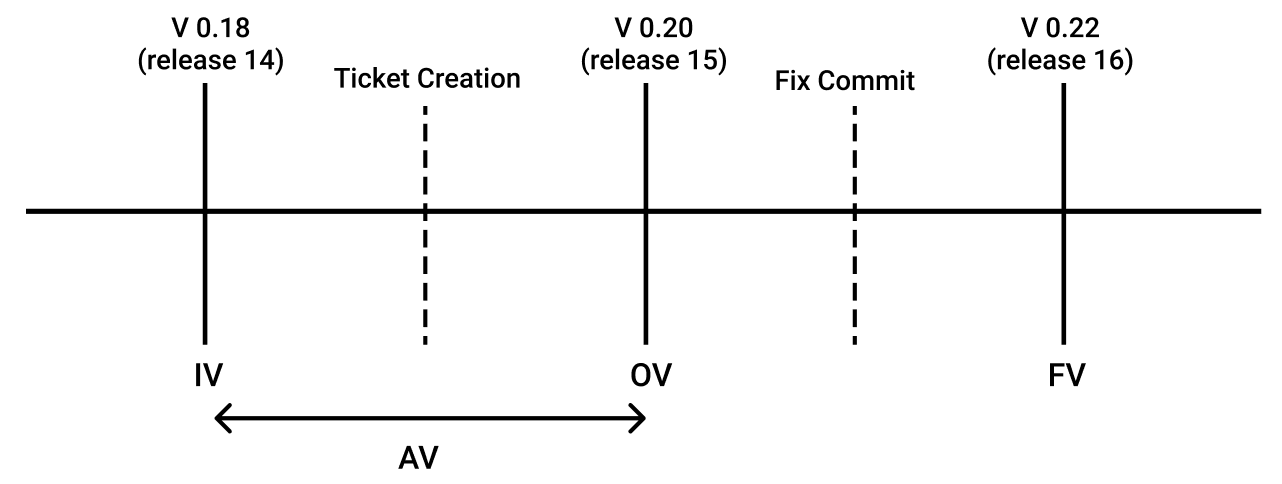}
  \vspace{-1mm}
    \caption{Example of the life-cycle of a defect: Introduction Version (IV), Opening Version (OV), Fixed Version (FV), and Affected Versions (AV). Note, versions V0.19 and V0.21 were only "baselines" and not "user-intended" versions and, hence, were excluded}
   \label{fig:example}
   \vspace{-3mm}
\end{figure}

Figure \ref{fig:example} illustrates the key terms while using the defect QPID-4462\footnote{https://issues.apache.org/jira/browse/QPID-4462} as an example. The defect is first injected in the code at the {\em Introducing Version} (\textbf{IV}), i.e., the V0.18 version in Figure~\ref{fig:example}. Afterwards, a failure is observed and a defect report is created to describe the defect. We refer to the version related to the creation of the defect report as the {\em Opening Version} (\textbf{OV}), i.e., the V0.20 version in Figure \ref{fig:example}. Next, in a given future version, the defect is fixed by changes performed in one or more classes. We refer to the version related to the fix of the defect as the {\em Fixing Version} (\textbf{FV}), i.e., the V0.22 version in Figure~\ref{fig:example}. An AV is any  version in which the defect could have caused a failure, i.e., any version {\em affected} by the defect. Thus, the AVs in our example are those in the range [IV, FV), i.e., the V0.18 and V0.20 versions in Figure~\ref{fig:example}. The V0.22 version is not an AV since it contains the fix and is not affected by the defect. 

The OV is available in all defect reports as it is generated by the issue tracker at the creation of the report. The FV is available in defect reports where developers have mentioned the defect report ID in the log of the commit that fixes the defect. For example, commit \textit{732ab160852f943cd847646861dd48370dd23ff3} is the last commit including \textit{[QPID-4462]} in its log. Since this commit was performed at \textit{2013-03-31T21:51:49+00:00}, we can infer that it has been performed between versions V0.20 and V0.22. 

Our intuition, that defects have a stable life cycle in terms of the proportion of the number of versions required to discover and to fix a defect, actually means that FV-OV is proportional to FV-IV.

\subsection{RQ1: \rqone}

\subsubsection{Design} 
\label{sec:DesRQ1}
In this section we report the design and result of our first research question.
\\
\textbf{Dependent variables}

 Our dependent variable is  the percentage of {\em available \& consistent} AVs. An AV is {\em available} if it is provided in the report related to a defect (i.e., the defect report). An AV information is  {\em consistent} when the earliest AV occurs before the OV. The rationale is that the defect must have affected a version that occurred at least at the moment when the defect report had been created. That is, a defect cannot have been injected {\em after} the related failure had been observed. 

\textbf{Measurement Procedure}
\\
To measure the availability and consistency of AVs, we follow the following nine steps:

\begin{enumerate}
\item We retrieve the JIRA and Git URL of all existing Apache projects.\footnote{https://people.apache.org/phonebook.html} We focused on Apache projects rather than GitHub projects because Apache projects have a higher quality of defect annotation and are unlikely to be toy projects \cite{Munaiah:2017:CGE:3147777.3147808}. Finally, Apache projects use JIRA as their issue tracker, which allows us to study the consistency and availability of AV information.
\item We filter out projects which are not tracked in JIRA or not versioned in Git. This leads to 212 projects.
\item As recently done by Borg et al. \cite{Borg2019}, for each project, we count the number of issue reports by performing the following query to the JIRA repository:
\textit{Type ==  ``defect'' AND (status == ``Closed'' OR status ==``Resolved'') AND Resolution ==``Fixed''}. This gave us a total of about 235,000 defects.
\item We exclude issue reports not having a related Git commit fixing it.
\item We exclude defects that are not post-release. Post-release defects are also known in the industry as production defects, i.e., defects that caused failures observed by users. Thus, a defect that is injected and fixed in the same version is not a post-release defect. For brevity, in the remainder of this paper, we refer to post-release defects simply as defects. After steps 4 and 5, we are left with a total of 125,000 defects.
\item For each defect report, we check its AV \textbf{availability}, i.e., the presence of the AV field, by performing the following query to the JIRA repository: Affect Version $\neq$ ``Null''. Thus, each issue report is tagged as available or unavailable.
 \item For each defect report, we check its AV \textbf{consistency}, i.e., if IV $\le$ OV.
 \item For each project, we compute the percentage of  unreliable and of unavailable AV.
 \item Finally, we observe the distribution of 212 Apache projects in terms of percentages of defects having available, and available \& consistent AV.
\end{enumerate}

\subsubsection{RQ1: Results}
\label{sec:ResRQ1}

\begin{figure}
  \includegraphics[width=0.6\columnwidth]{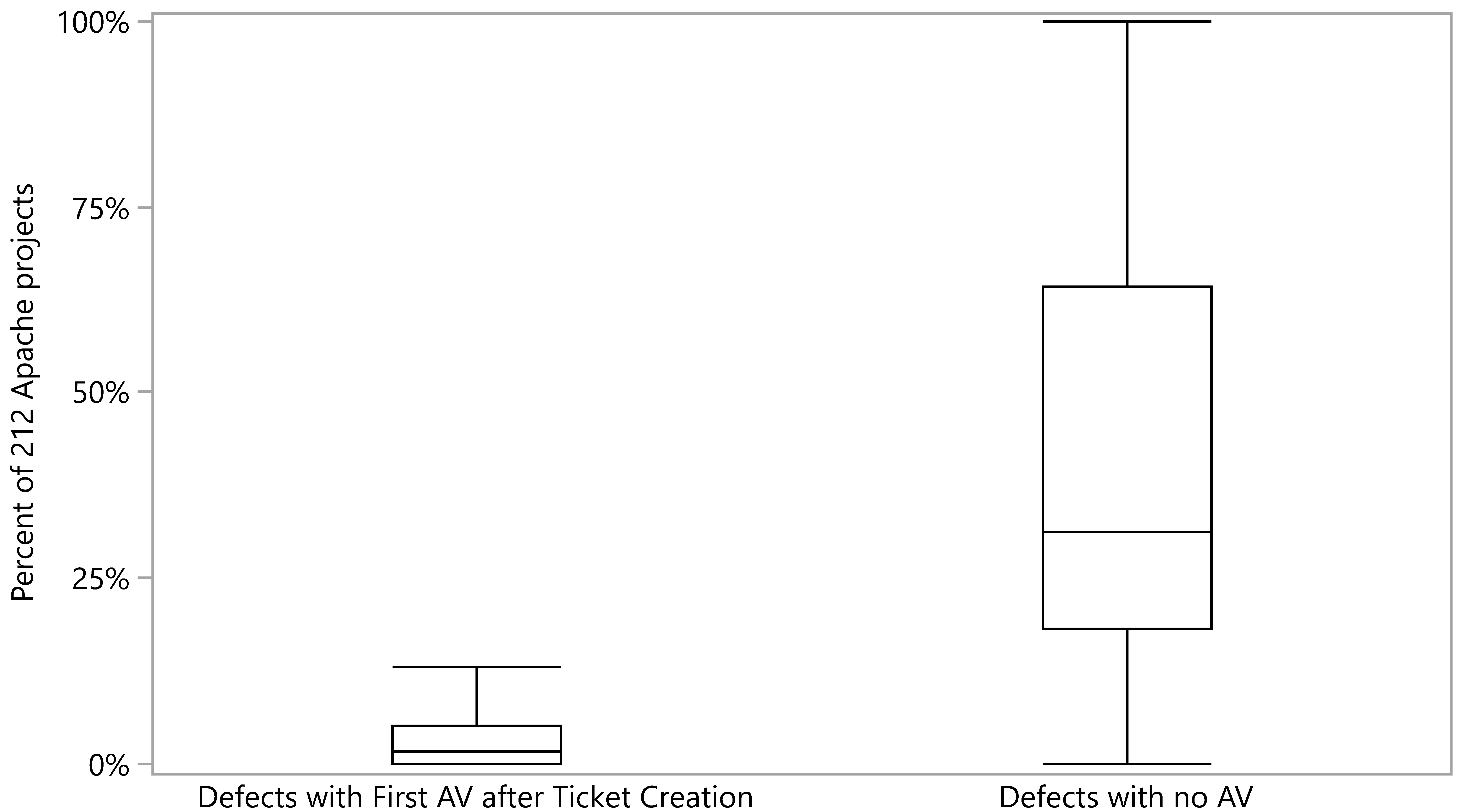}
  \vspace{-1mm}
    \caption{Distribution of 212 Apache projects having a specific proportion of defect reports with an unreliable AV (left side) or without the AV (right side).}
   \label{fig:RQ1v1}
   \vspace{-3mm}
\end{figure}

Figure \ref{fig:RQ1v1} reports the distribution of 212 Apache projects having a specific proportion of defects with an unreliable AV (left side) or without the AV (right side). According to Figure \ref{fig:RQ1v1}, most of the projects have more than 25\% of defect reports without any AV. 
We also measured the total number of closed defect reports linked with git commits in the 212 Apache projects, which resulted to be 125,860. Of these, 63,539 defect reports (51\%) resulted in not having or having inconsistent AVs. Thus, we can claim that \textbf{\em in most of defect reports, we cannot use the AVs and,  hence, we often need an automated method for retrieving AVs.} 
 
\subsection{RQ2: \rqtwo}

\subsubsection{Design} 
\label{sec:DesRQ2}
In this section, we report the design and result of our second research question.
Since AVs are in the range [IV, FV), and since we always know FV, retrieving the AVs of a defect actually translates into estimating its IV.
One approach to estimate the IV of a defect is to employ the SZZ algorithm. The oldest defect-introducing commit produced by SZZ can be considered as the IV, whereas all other defect-introducing commits can be used to label the consecutive versions before the defect-fixing commit as other AVs (the defect-fixing commit itself is not considered in the labeling process, of course). However, existing researches have highlighted substantial limitations of the SZZ approach~\cite{CostaMSKCH17,DBLP:journals/infsof/Rodriguez-Perez18,Rodriguez-Perez:2018:BDO:3239235.3267436}.

We investigate the following null hypothesis in this RQ: 

{\em $\bullet H_{10}$: different methods obtain the same accuracy for labeling AVs.}

\textbf{Independent variables}

Our independent variable is the method used to retrieve the AV, i.e., to label a version as affected or not by a specific defect.
In this work, we present three variants of a new approach called {\em Proportion} to label affected versions. The Proportion approach assumes a stable proportion ($P$), among defects of the same project, between the number of affected versions between IV and FV, and the number of versions between OV and FV. The rationale is that the life-cycle might be consistent among defects of the same projects. Thus, in some projects, defects require a number of versions to be found and another number to be fixed. Our intuition is that the proportion among these numbers is somehow stable across defects of the same project. Of course, defects of the same projects may vary and, hence, we do not expect this method to be perfectly accurate. Since FV and OV are known for every defect, the idea is to compute $P$ on previous defects, and then use it for defect reports where AVs are not available nor consistent. Thus, we define $P$ as $(FV - IV)/ (FV - OV)$. Therefore, we can calculate the IV as $FV - (FV - OV) * P$. Among the possible ways to use Proportion we propose the following three methods:
    \begin{itemize}
     \item \textbf{Proportion\_Incremental:} It consists of computing P of the current defect as the average P on past defects of the same project. This approach has the advantage of using the highest amount of information, within the same project of the current defect, available at a given point in time. The assumptions of this approach are that 1) P does not change over time within the same project and that 2) there is enough information on previous defects of the project, i.e., there are enough previous defects to represent the true project average P. Specifically, in this method, we ordered the defects by fix date. For each version R within a project, we used the average $P$ among defects fixed in versions 1 to R-1. Using the example in Figure \ref{fig:example}, the $P\_Increment$, computed as the average P among defects in versions 1 to 15, is $1.7775$. Therefore, $IV = 16 - (16 - 15) * 1.7775$, which is $14.2225$. Hence, this method would correctly identify 0.20 as affected (true positive), but incorrectly classify 0.18 as not affected (false negative).
     \item \textbf{Proportion\_ColdStart:} It consists of computing P of the current defect as the average P on defects of other projects. This approach has the advantage of using the knowledge related to other projects and it is particularly useful for the first defects in a project, i.e., when there are not enough previous defects to represent the true project average P. The assumption of this approach is that P is stable across defects of different projects. Specifically, for each studied project, we compute the average $P$ across all defects within the project. We label each of these projects as $P\_PROJECT$ where $PROJECT$ is the project's ID. Next, for each project, we take the median of the $P\_PROJECT$ values among all other projects to use as the $P\_ColdStart$. Using the example in Figure \ref{fig:example}, the indexes of the 0.18, 0.20, and 0.22 versions are 14, 15, and 16, respectively. The $P\_ColdStart$ computes as the median of the other $P\_PROJECT$ proportions, as $1.8089$. Therefore, $IV = 16 - (16 - 15) * 1.8089$ which is $14.1911$. Hence, this method would correctly identify 0.20 as affected (true positive), but incorrectly classify 0.18 as not affected (false negative).
    \item \textbf{Proportion\_MovingWindow:} : It consists of computing P of the current defect as the average P on a limited number of recent defects of the same project. This approach has the advantage of relaxing the assumption that P does not change over time within the same project as we limit the amount of time where this assumption needs to hold, as opposed to Proportion\_Incremental where the assumption needs to hold for the entire length of the project. The assumptions of this approach are that 1) the length of window is big enough to contain sufficient previous defects to represent the true project average P and 2) the window is small enough to exclude defects that are different to the current one. Therefore, the hard part in implementing the Proportion\_MovingWindow method is in defining the right length of the window. Specifically, we ordered the defects by their fix date. For each defect within a project, we used the average P among the last 1\% of fixed defects. We chose a 1\% moving window length as a tradeoff between the ability to react to changes in the data and the resilience to outliers. Using the example in Figure \ref{fig:example}, the $P\_MovingWindow$ is computed as the average P among the last 1\% of defects. There are 1,192 defects in the project of Figure~\ref{fig:example}. Therefore, there are around 12 defects at the 1\% of defects. The average P among the last 12 fixed defects is $2.167$. Therefore, $IV = 16 - (16 - 15) * 2.167$ which is $13.833$. Hence, this method would correctly identify 0.18 and 0.20 as affected (true positive), giving 100\% accuracy for this defect.
    \end{itemize}

Moreover, a simplistic way to apply Proportion is to assume IV equals to OV. Therefore, we propose the following simplistic method:
\begin{itemize}

\item \textbf{Simple}: It simply assumes that the IV corresponds to OV. The rationale is that, by definition, all versions from OV to FV (not including FV) are AV. However, versions before OV can also be AV. Therefore, we expect this heuristic to achieve a 100\% Precision but a lower Recall. Specifically, this heuristic would identify 0.20 as IV in Figure \ref{fig:example}. Therefore, it would miss 0.18 (false negative) and would correctly identify 0.20 (true positives) as AVs. 
\end{itemize}

Furthermore, we considered as baseline the well known SZZ approach. As previously discussed in Section \ref{sec:RelatedWork}, SZZ is an algorithm that, given a fix commit, determines the possible defect-introducing commits. In our methods, we assume the oldest defect-introducing commit to be the IV. Specifically,  among the possible ways to use SZZ, we considered the following methods:
    \begin{itemize}  
      \item \textbf{SZZ\_Basic}: We use the SZZ algorithm \cite{Sliwerski:2005:CIF:1082983.1083147} to determine when the defect has been introduced, and we assume as AVs all versions between the IV and the FV (not including FV). In the example in Figure \ref{fig:example}, SZZ\_B identified three defect-introducing commits with the following dates: 2012-05-19T08:54:25, 2012-10-06T05:38:51, 2012-11-05T10:03:36. Among these, 2012-05-19T08:54:25 is the oldest date, which falls into version 0.18 labeled as the IV. Therefore, the AVs are  0.18 and 0.20. Versions 0.18 and 0.20 were correctly identified as affected (true positives), and therefore, this method receives 100\% accuracy for this defect.
    \item\textbf{SZZ\_U}: We rely on an open implementation of SZZ by Borg et al. \cite{abs-1903-01742} and we set the depth to one. This SZZ implementation does not discard cosmetic changes (since it supports all programming languages). However SZZ\_U uses Jaccard distances to map moving lines. In the example in Figure \ref{fig:example}, SZZ\_U identified one defect-introducing commit dated 2012-05-18T20:54:25, which falls into version 0.16 labeled as the IV. Therefore, the AVs are  0.16, 0.18, and 0.20. Versions 0.18 and 0.20 were correctly identified as affected (true positives) and version 0.16 was incorrectly identified as affected (false positives).
    \item\textbf{SZZ\_RA}: We use a refactoring-aware SZZ algorithm implemented by Da Costa \cite{CostaMSKCH17}. This algorithm tracks defect-introducing commits and filters out refactoring operations. However, this implementation only analyzes java files, so the defect-introducing commits for non-java files are determined by SZZ\_U. In the example in Fig. \ref{fig:example}, SZZ\_RA identified one defect-introducing commit dated 2012-05-18T16:54:25 which falls into version 0.16 labeled as the IV. Therefore, the AVs are  0.16, 0.18, and 0.20. Versions 0.18 and 0.20 were correctly identified as affected (true positives) and version 0.16 was incorrectly identified as affected (false positives).
\end{itemize}

Finally, instead of using the SZZ\_X methods as is, we improved them by using the information provided by the Simple method. Specifically, we created methods \textbf{SZZ\_X+} by merging each {\em SZZ\_X} with the {\em Simple} method: a version is affected if SZZ\_X labeled it as affected or Simple labeled it as affected. Hence, we are merging the defects' life cycle information with the SZZ based method. The rationale is that if Simple labels a version as affected, then the version is actually affected by definition. To illustrate how this works, we will use a new example, WICKET-4071\footnote{https://issues.apache.org/jira/browse/WICKET-4071}. The AVs indicated on this defect report are: 1.4.6, 1.4.7, 1.4.8, 1.4.19, 1.4.10, and 1.5-M1. The OV is 1.4.8 and FV is 1.5-M1. Simple would classify versions 1.4.8, 1.4.19, 1.4.10, and 1.5-M1 as affected (true positives) and would miss versions 1.4.6 and 1.4.7 (false negatives). SZZ\_B would classify 1.4.10 and 1.5-M1 as affected (true positives) and miss versions 1.4.6, 1.4.7, 1.4.8, and 1.4.19 (false negatives). However, SZZ\_B+ would classify versions 1.4.8, 1.4.19, 1.4.10, and 1.5-M1 as affected (true positives) and would miss versions 1.4.6 and 1.4.7 (false negatives).

In conclusion, in this paper we analyze the accuracy of three already known methods (SZZ\_X) and seven new methods (Proportion\_X, SZZ\_X+, and Simple)

\textbf{Dependent variables}

Our dependent variable is the accuracy for labeling versions of a project as affected, or not, by a defect. We use the following set of metrics:

\begin{itemize}
    \item True Positive(TP): The version is actually  affected and is labeled as affected.
    \item False Negative(FN): The version is actually  affected and is labeled as non-affected.
    \item True Negative(TN): The version is actually  non-affected and is labeled as non-affected.
    \item False Positive(FP):  The version is actually  non-affected and is labeled as affected.
    \item \textbf{Precision} : $\frac{TP}{TP + FP}$ 
    \item \textbf{Recall} : $\frac{TP}{TP + FN}$  
    \item \textbf{F1} : $\frac{2 * Precision * Recall}{Precision + Recall}$
    \item Cohen's \textbf{Kappa} : A statistic that assesses the classifier's performance against random guessing \cite{cohen1960coefficient}. $Kappa = \frac{Observed - Expected}{1 - Expected} $ where 
    \begin{itemize}
        \item Observed: The proportionate agreement. $\frac{TP + TN}{TP + TN + FP + FN}$
        \item Expected: The probability of random agreement. $P_{Yes} + P_{No}$ where 
        \begin{itemize}
            \item $P_{Yes}$: Probability of positive agreement. \\ $\frac{TP + FP}{TP + TN + FP + FN} * \frac{TP + FN}{TP + TN + FP + FN}$
            \item $P_{No}$: Probability of negative agreement. \\ $\frac{TN + FP}{TP + TN + FP + FN} * \frac{TN + FN}{TP + TN + FP + FN}$
        \end{itemize}
    \end{itemize}{}
    \item \textbf{Matthews} Correlation Coefficient : $\frac{TP * TN - FP * FN}{\sqrt{(TP+FP)(TP+FN)(TN+FP)(TN+FN)}}$
\end{itemize}
Since we have binary classifications that are thresholds independent, we do not use Area Under the Receiver Operating Characteristic metric.

\textbf{Measurement procedure}

\begin{figure}[H]
\begin{center}
   \includegraphics[width=1.0\columnwidth]{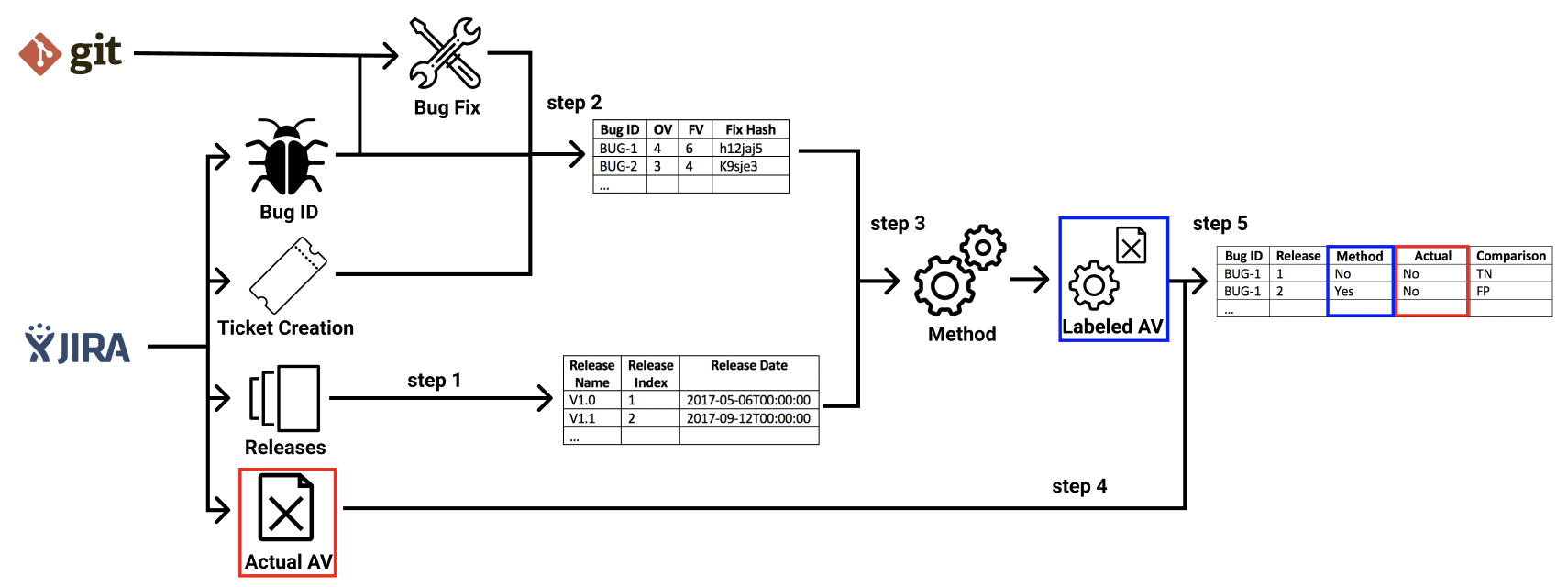}
  \vspace{-1mm}
    \caption{The process to measure the accuracy of methods in labeling affected versions.}
   \label{fig:rq2process}
   \vspace{-3mm} 
\end{center}
\end{figure}

We began by selecting the projects with the highest proportion of usable (i.e., available and consistent) AVs. We selected projects with at least 100 defects that were linked with git and contained available and consistent AVs. Then, we filtered out projects with less than 6 versions. Lastly, we filtered out projects where the percent of available and consistent AVs are less than 50\%. This left us with 76 projects. For each project, we followed the steps below. See Figure \ref{fig:rq2process} for an overview of this process.
\begin{enumerate}
\item We retrieved the versions of the project and their release dates from JIRA. We numbered these versions beginning with the oldest version as version 1.
\item We used the defects, of which their reports provided available and consistent AVs in RQ1. For each defect, we determined the IV (i.e., the version of the first AV labeled by JIRA), OV (i.e., the version of the ticket creation), FV (i.e., the fix version), and the fix commit hash by Git. We ordered the defects by fix date.
\item For each defect, we labeled versions 1 to FV as affected or not by each of the following methods:
    \begin{enumerate}
            \item \textbf{Simple:}
            \begin{enumerate}
                \item We set IV equal to OV.
                \item For each defect, we label each version before the IV as not affected. We label each version from the IV to the FV as affected. The FV is labeled not affected.  
            \end{enumerate}
        \item \textbf{SZZ:}
            \begin{enumerate}
                \item We ran each SZZ implementation on the project by supplying the Git directory and a list of defects and their fix commit. 
                \item For each defect, SZZ outputs all possible defect-introducing commits. We compute the corresponding version for each defect-introducing commit. We chose the oldest version to be the IV.
                \item For each defect, we label each version before the IV as not affected. We label each version from the IV to the FV as affected. The FV is labeled not affected.  
            \end{enumerate}
        \item \textbf{Proportion\_ColdStart:}
            \begin{enumerate}
                \item We computed the average P across the project's defects, i.e., $P = (FV-AV)/(FV-OV)$. If FV equals OV, then $FV-OV$ is set to one to avoid divide by zero cases.
                \item We computed the $P\_ColdStart$, i.e., the median P of all other projects.
                \item For each defect, we computed the IV as $IV= (FV-OV) * P\_ColdStart$. If FV equals OV, the IV equals FV. However, recall we excluded defects that were not post-release. Therefore, we set $FV - OV$ equal to 1 to assure IV is not equal to FV.
                \item For each defect, we label each version before the IV as not affected. We label each version from the IV to the FV as affected. The FV is labeled not affected.  
            \end{enumerate}
        \item \textbf{Proportion\_Increment:}
            \begin{enumerate}
                \item For each version R, we computed $P\_Increment$ as the average P among defects fixed in versions 1 to R-1.
                \item We used the P\_ColdStart for P\_Increment values containing less than 5 defects in the average.
                \item For each defect in each version, we computed the IV as $IV = (FV-OV) * P\_Increment$. If FV equals OV, the IV equals FV. However, recall we excluded defects that were not post-release. Therefore, we set $FV - OV$ equal to 1 to assure IV is not equal to FV.
                \item For each defect, we label each version before the IV as not affected. We label each version from the IV to the FV as affected. The FV is labeled not affected.  
            \end{enumerate}
        \item \textbf{Proportion\_MovingWindow:}
            \begin{enumerate}
                \item For each defect, we computed P\_MovingWindow as the average P among the last 1\% of defects. The defects are ordered by their fix date.
                \item We used the P\_ColdStart for P\_MovingWindow values containing less than 1\% of defects in the average.
                \item For each defect, we computed the IV as $IV = (FV-OV) * P\_MovingWindow$. If FV equals OV, the IV equals FV. However, we excluded defects that were not post-release. Therefore, we set $FV - OV$ equal to 1 to assure IV is not equal to FV.
                \item For each defect, we label each version before the IV as not affected. We label each version from the IV to the FV as affected. The FV is labeled not affected.  
            \end{enumerate}
        \item \textbf{+:}
            \begin{enumerate}
                \item For each SZZ method, we combined it with Simple. For each defect, we labeled each version as affected if SZZ\_X or Simple labeled the version as affected.  
            \end{enumerate}
    \end{enumerate}
    \item We determined the observed/actual AV by looking at JIRA values. We label each version before the IV, as labeled by JIRA developers, as not affected. We label each version from the IV to the FV, as labeled by JIRA developers, as affected. The FV, as labeled by JIRA developers, is labeled not affected.
    \item For each method, we compared the classification to the actual classification and computed the TP, TN, FP, FN, Precision, Recall, F1, Matthews, and Kappa across the project's version-defect pairs.
\end{enumerate}

\textbf{Hypothesis testing.}
To test hypothesis $H_{10}$, we used the Kruskal--Wallis test~\cite{Kruskal:ASA:1952}, which is a non-parametric test (i.e., a normally distributed data is not required) to check whether three or more distributions are significantly different. This test is necessary because we compare 10 distributions of values (i.e., one for each studied method). For example, we compare whether our 10 studied methods for estimating AVs significantly differ in terms of precision or recall. We chose the non-parametric Kruskal--Wallis because our performance metrics (e.g., precision or recall) do not follow a normal distribution (as we noted when performing Shapiro--Wilk tests~\cite{Shapiro:Biometrika:1965}). Therefore, our approach is compliant to the suggestion to avoid using ScottKnottESD in case of not normal distributions \cite{DBLP:journals/tse/Herbold17}.

Although Kruskal--Wallis provides an indication of whether a significant difference between distributions exists, it does not indicate which specific pairs of distributions are significantly different between each other. For this purpose, we perform a follow-up Dunn test~\cite{Dunn:Technometrics:1964}, which is a {\em post-hoc} test to indicate which specific pairs of distributions significantly differ between each other. To account for the chance of errors due to multiple comparisons, we perform a Holm-Bonferroni correction of our $p-values$~\cite{Holm:SJS:1979}. 

\subsubsection{RQ2: Results}
\label{sec:ResRQ2}
\begin{figure}
  \includegraphics[width=0.9\columnwidth]{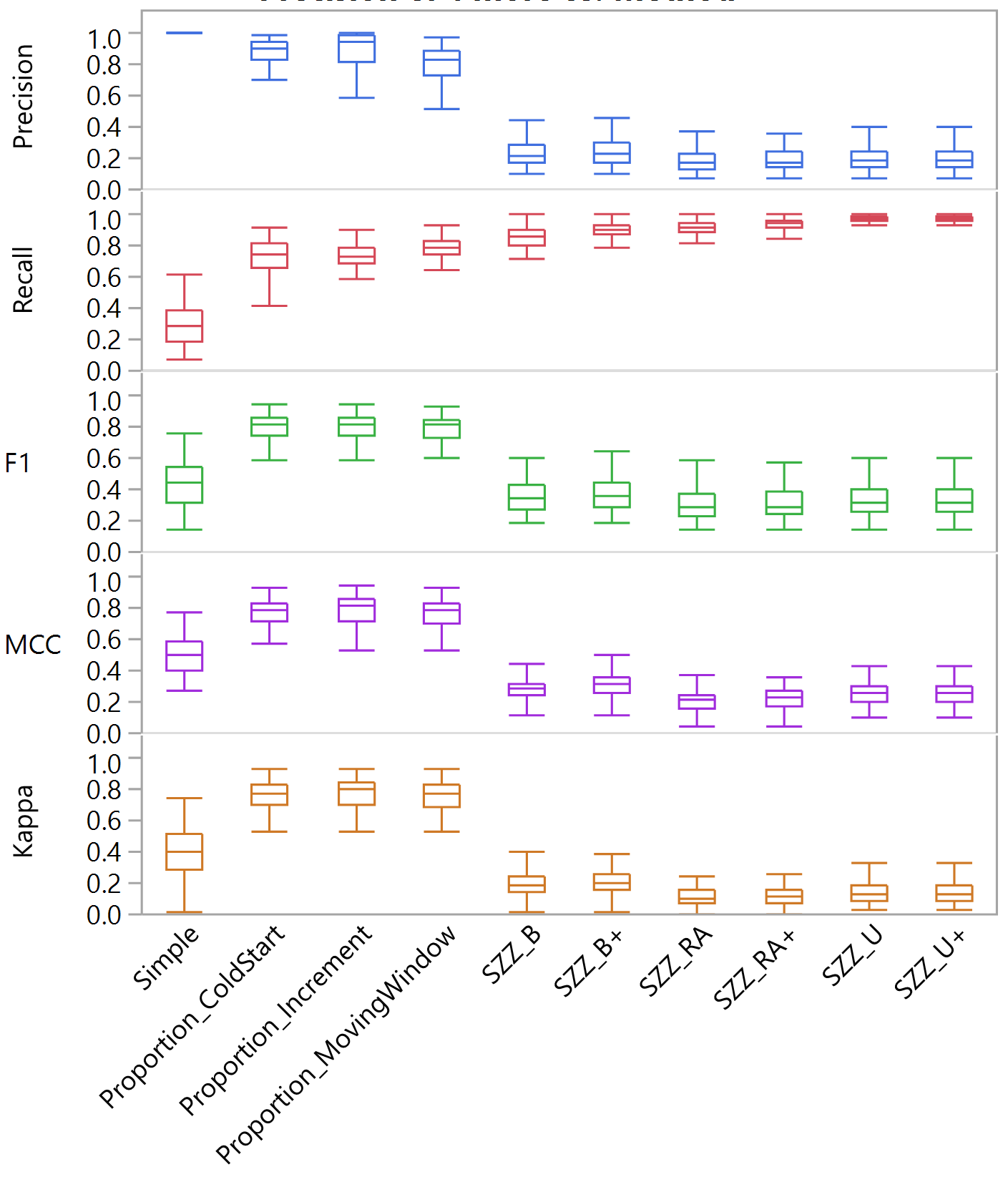}
  \vspace{-1mm}
    \caption{Distribution, across 76 Apache projects, of Precision, Recall, F1, MCC, and Kappa, of different methods in labeling AV.}
   \label{fig:RQ2result}
   \vspace{-3mm}
\end{figure}

Figure \ref{fig:RQ2result} reports the distribution, across 76 Apache projects, of Precision, Recall, F1, MCC, and Kappa, of different methods in labeling AV.
According to Figure \ref{fig:RQ2result}:
\begin{itemize}
    \item All the Proportion methods have a higher Precision and composite accuracy (F1, MCC, and Kappa) than all SZZ methods. 
    \item Simple has a higher Precision and composite accuracy (F1, MCC, and Kappa) than all SZZ methods. 
    \item SZZ\_U has the highest Recall than all other methods.
    \item SZZ\_B+ has the highest Precision and the highest composite accuracy (F1, MCC, and Kappa) than any other SZZ method.
    \item The method with the highest precision is Simple. This is true by definition. 
    \item There is no single dominant method among the Proportion methods. For instance, Proportion\_Increment provides the highest Precision, F1 and Kappa and it dominates Proportion\_ColdStart. Proportion\_MovingWindow provides the highest Recall (among Proportion methods) and MCC.
\end{itemize}   

\begin{table}
    \centering
    \begin{tabular}{lrr}
    \hline
         {\bfseries Method} & {\bfseries Mean Precision} & {\bfseries Dunn's test Rank}  \\
         \hline
         \hline
         Simple & 1.0 & 1\\
         Proportion Increment & 0.90 & 2\\
         Proportion Cold Start & 0.89 & 2\\
         Proportion Moving Window & 0.81 & 3\\
         SZZ\_B+ & 0.26 & 4\\
         SZZ\_B & 0.25 & 4\\
         SZZ\_U+ & 0.22 & 5\\
         SZZ\_U & 0.22 & 5\\
         SZZ\_RA+ & 0.21 & 5\\
         SZZ\_RA & 0.20 & 5\\
         \hline
         \hline
    \end{tabular}
    \caption{Dunn's results for the {\em precision} values of hypothesis $H_{10}$}.  
    \label{tab:rq2:precision_hypothesis_tests}
\end{table}

\begin{table}
    \centering
    \begin{tabular}{lrrl}
    \hline
         {\bfseries Method} & {\bfseries Avg. Recall} & {\bfseries Dunn's test Rank} & {\bfseries Comment}  \\
         \hline
         \hline
         SZZ\_U+ & 0.97 & 1 &\\
         SZZ\_U & 0.97 & 1 &\\
         SZZ\_RA+ & 0.94 & 1.5 & Significantly lower than SZZ\_U+\\
         SZZ\_RA & 0.91 & 2 &\\
         SZZ\_B+ & 0.89 & 2 &\\
         SZZ\_B & 0.85 & 2.5 & Significantly lower than SZZ\_RA\\
         Proportion Moving Window & 0.78 & 3 &\\
         Proportion Increment & 0.73 & 3 &\\
         Proportion Cold Start & 0.72 & 3 &\\
         Simple & 0.30 & 4\\
         \hline
         \hline
    \end{tabular}
    \caption{Dunn's test results for the {\em recall} values of hypothesis $H_{10}$. We increased the rank by a half whenever a method is significantly different from only one other method within the group.}.  
    \label{tab:rq2:recall_hypothesis_tests}
\end{table}

\begin{table}
    \centering
    \begin{tabular}{lrrl}
    \hline
         {\bfseries Method} & {\bfseries Avg. F1} & {\bfseries Dunn's test Rank} & {\bfseries Comments}  \\
         \hline
         \hline
         Proportion Increment & 0.80 & 1 &\\
         Proportion Moving Window & 0.80 & 1 &\\
         Proportion Cold Start & 0.79 & 1 &\\
         Simple & 0.44 & 2 &\\
         SZZ\_B+ & 0.39 & 2 &\\
         SZZ\_B & 0.37 & 2 &\\
         SZZ\_U+ & 0.35 & 2 &\\
         SZZ\_U & 0.35 & 2 &\\
         SZZ\_RA+ & 0.32 & 2 &\\
         SZZ\_RA & 0.32 & 2.5 & Significantly lower than Simple\\
         \hline
         \hline
    \end{tabular}
    \caption{Dunn's test results for the {\em f1} values of hypothesis $H_{10}$. We increased the rank by a half whenever a method is significantly different from only one other method within the group.}  
    \label{tab:rq2:f1_hypothesis_tests}
\end{table}

Our statistical results on the 76 Apache projects reveal that the differences between our studied methods in terms of the accuracy metrics are statistically significant for $H_{10}$ (i.e., our Kruskall--Wallis and Dunn's tests yielded $p-values < 0.05$). Therefore, our results reveal that the proportional methods have significantly better accuracy values compared to the studied SZZ based methods.
Tables~\ref{tab:rq2:precision_hypothesis_tests},~\ref{tab:rq2:recall_hypothesis_tests},~and~\ref{tab:rq2:f1_hypothesis_tests} show the results of the Dunn's tests for the precision, recall, and F1 metrics, respectively. For each table, we show (i) the methods, (ii) the {\em mean} accuracy, and (iii) the {\em rank} of the method. The {\em rank} is based on whether the Dunn's test provided a significant $p-value$ for a given comparison. For example, in Table~\ref{tab:rq2:precision_hypothesis_tests}, the {\em Simple} approach is placed at the top rank because (a) it has the highest mean value and (b) the Dunn's tests yielded significant $p-values$ for every pair-wise comparison between the {\em Simple} method and the other methods. On the other hand, the {\em Proportion Increment} and {\em Proportion Cold Start} methods are placed at the second rank because the Dunn's test revealed that these two specific methods do not have statistically different precisions. However, both the {\em Proportion Increment} and {\em Proportion Cold Start} methods have significant $p-values$ when compared to the other methods that fell in lower ranks.

According to our observations, the {\em Proportion Increment}, {\em Proportion Moving Window}, and {\em Proportion Cold Start} methods are all in the 1st rank in terms of F1. However, the {\em Proportion Increment} and {\em Proportion Cold Start} methods yield significantly better precision values (i.e., $p-values < 0.05$) compared to the {\em Proportion Moving Window} method (see Table~\ref{tab:rq2:precision_hypothesis_tests}).
Interestingly, even the {\em Simple} method (which fell in the 2nd rank, see Table~\ref{tab:rq2:f1_hypothesis_tests}) significantly outperforms the SZZ\_RA method.

\subsection{RQ3: \rqthree}
\subsubsection{Design} 
\label{sec:DesRQ3}

We propose the following hypothesis for this RQ:

{\em $\bullet H_{20}$: different methods obtain the same accuracy for labeling classes.}

Our empirical procedure wa inspired by \cite{DBLP:journals/tse/CostaMSKCH17} and it is detailed in the following subsections.

\textbf{Independent variables}

The independent variable is represented by the same methods shown in RQ2. However, in this research question, the retrieved AVs (as performed in RQ2) is used to label classes as defective or not. 

\textbf{Dependent variables}

\begin{figure}[H]
\begin{center}
  \includegraphics[width=1.0\columnwidth]{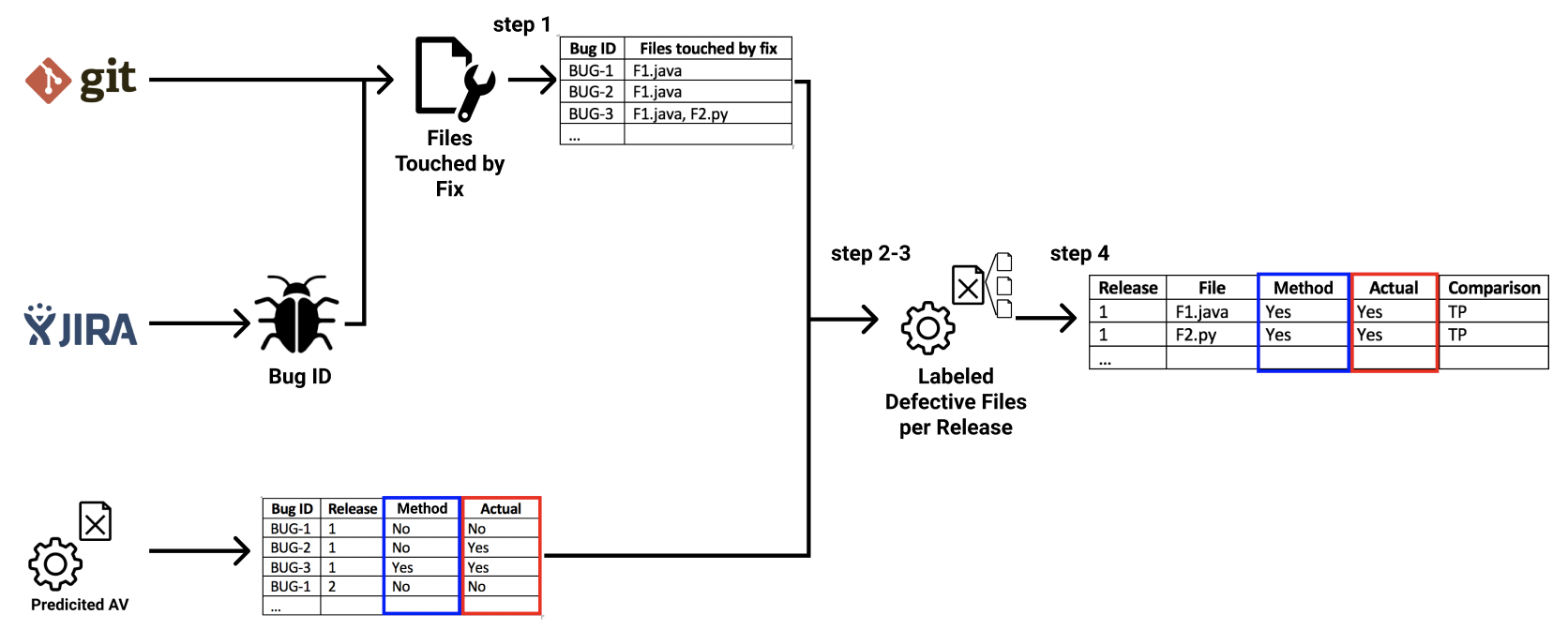}
  \vspace{-1mm}
    \caption{The process to measure the accuracy of methods in labeling defective classes.}
   \label{fig:rq3process}
   \vspace{-3mm}
  \end{center}
\end{figure}

The dependent variables are the same accuracy metrics presented in RQ2, with the only difference that the unit upon which the accuracy is computed is the defectiveness of a class in a version. 
If at least one defect impacts the version-class pair, then the version-class pair is labeled as defective. This is demonstrated in Figure \ref{fig:rq3process} where F1.java is deemed defective because it was touched by the fix for defect-3 in version 1 (i.e., at least one defect-fix touched F1.java in version 1).

In order to better explain the difference between RQ3 and RQ2, let's consider the case of methods A, B, and C, and a class that was affected by three defects in a certain version. Suppose that A is able to identify that the class was affected by one defect, B, by three defects, and C, by 4 defects. In this example, all three methods correctly identify the class in the version as defective and, therefore, all three methods result with perfect accuracy. However, for the purpose of RQ2, method B has a higher accuracy than methods A and C.
The following metrics have been redefined for this RQ:

\begin{itemize}
    \item True Positive(TP): The class in a version is actually defective and is labeled as defective.
    \item False Negative(FN): The class in a version is actually  defective and is labeled as non-defective.
    \item True Negative(TN): The class in a version is actually non-defective and is labeled as non-defective.
    \item False Positive(FP):  The class in a version is actually non-defective and is labeled as defective.
\end{itemize}

\textbf{Measurement procedure}

Figure \ref{fig:rq3process} describes the process we use to label a class in a version as defective or not. The process is identical to what \citet{Yatish2019} coined as the realistic approach. The only difference that the AV is assumed to be unavailable and, hence, it is retrieved by using a certain proposed method (see RQ2). The process consists of three steps:

\begin{enumerate}
    \item For each defect in RQ2, we computed a list of classes touched by the fix commit.
    \item For each method in RQ2, we labeled each version-class pair as defective if the version of the pair was determined to be an AV of at least one defect in RQ2 and the  defect-fix commit of that defect involved the analyzed class. Otherwise, the version-class pair was labeled as not defective.
    \item We determined the observed/actual defectiveness of each version-class pair. To this end, we labeled each version-class pair as defective if the version of the pair was indicated as an AV of at least one defect by JIRA developers themselves, and the defect-fix commit of that defect touched the class. Otherwise, the version-class pair was labeled as not defective. To identify which commit is related to which defect we looked for the ticket's ID reported in the comment of the commit. For instance, consider the case of the defect ticket “QPID-4462”. All classes touched by all commits reporting the string “QPID-4462” are considered affected by defect “QPID-4462".
    \item For each proposed method, we compared its classifications to the observed/actual classification. Next, we computed the TP, FN, TN, FP, Precision, Recall, F1, Matthews, and Kappa metrics across the projects.
\end{enumerate}

\textbf{Hypothesis testing.}
To test hypothesis $H_{20}$, we use the same statistical machinery used in RQ2. We perform Kruskal-Wallis tests to check whether the distributions are significantly different. Next, we perform Dunn tests to identify the specific pairs of distributions that are significantly different between themselves.

\subsubsection{RQ3: Results}
\label{sec:ResRQ3}
\begin{figure}
  \includegraphics[width=0.9\columnwidth]{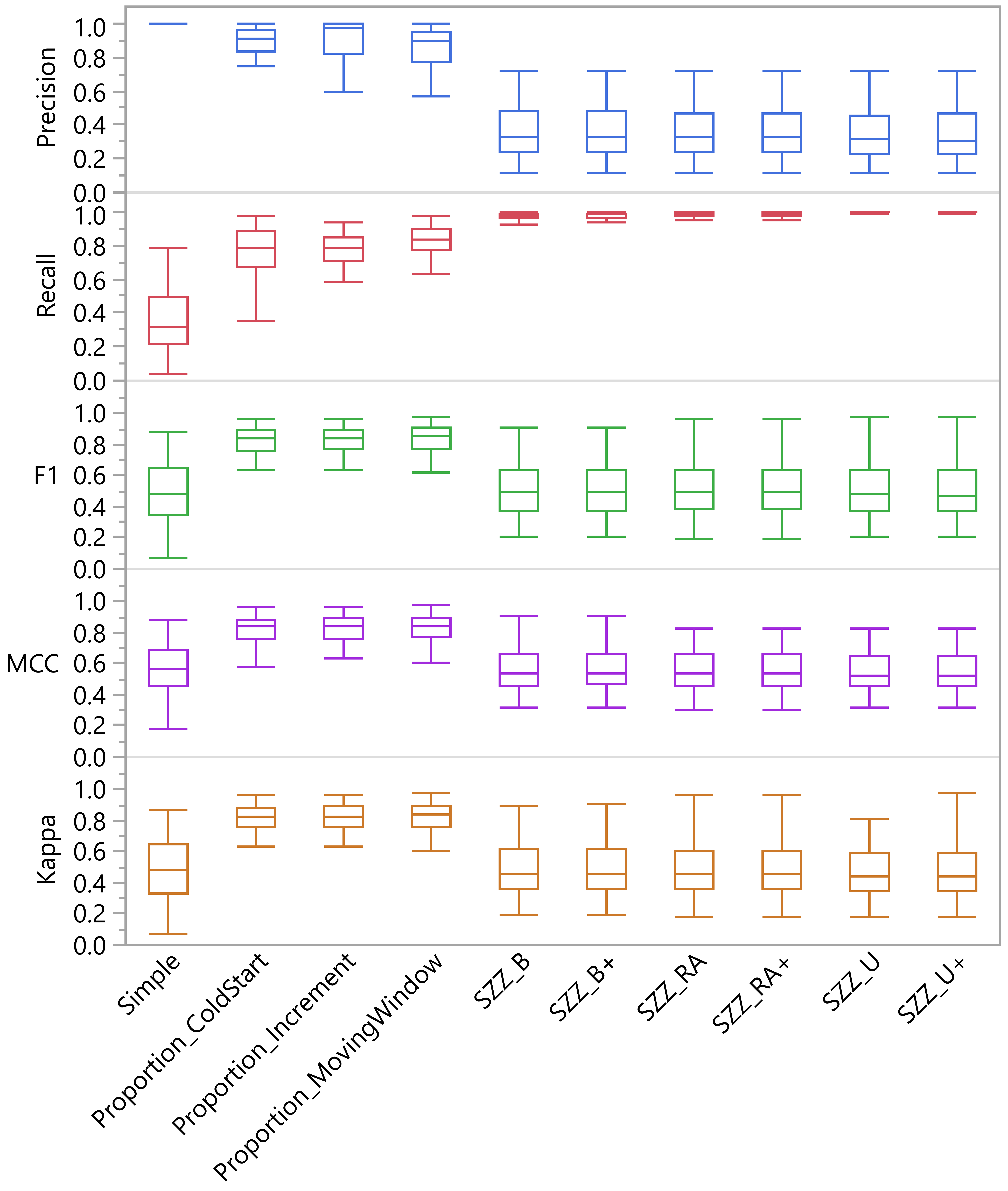}
  \vspace{-1mm}
    \caption{Distribution, across 76 Apache projects, of Precision, Recall, F1, MCC, and Kappa, of different methods in labeling defective classes.}
   \label{fig:RQ3result}
   \vspace{-3mm}
\end{figure}

 Figure \ref{fig:RQ3result} reports the distribution, across 76 Apache projects, of Precision, Recall, F1, MCC, and Kappa, of different methods for labeling defective classes. According to Figure \ref{fig:RQ3result}:
\begin{itemize}
    \item All the proportional methods have a higher Precision and composite accuracy (F1, MCC, and Kappa) compared to all SZZ methods. Therefore, we can claim that \textbf{labeling classes using defects' life cycle information is in overall and, in average, more accurate than the studied SZZ methods.}
    \item SZZ\_U has the highest Recall than all other methods.
    \item SZZ\_B+ has a highest Precision and lower Recall than any other SZZ method.
    \item SZZ\_B+ has a higher composite accuracy (F1, MCC, and Kappa) than Simple and any other SZZ method.
    \item \textbf{The Proportion\_MovingWindow method dominates all methods on all composite accuracy (F1, MCC, and Kappa)}. 
\end{itemize}   

\begin{table}
    \centering
    \begin{tabular}{lrrl}
    \hline
         {\bfseries Method} & {\bfseries Mean Precision} & {\bfseries Dunn's test} & {\bfseries Comments}  \\
         \hline
         \hline
         Simple & 1.0 & 1 &\\
         Proportion Increment & 0.91 & 1 &\\
         Proportion Cold Start & 0.90 & 1.5 & Significantly lower than Simple\\
         Proportion Moving Window & 0.86 & 1.5 & Significantly lower than Simple\\
         SZZ\_B+ & 0.38 & 2 &\\
         SZZ\_B & 0.37 & 2 &\\
         SZZ\_RA+ & 0.37 & 2 &\\
         SZZ\_RA & 0.37 & 2 &\\
         SZZ\_U+ & 0.36 & 2 &\\
         SZZ\_U & 0.36 & 2 &\\
         \hline
         \hline
    \end{tabular}
    \caption{Dunn's test results for the {\em precision} values of hypothesis $H_{20}$. We increased the rank by a half whenever a method is significantly different from only one other method within the group.}.  
    \label{tab:rq3:precision_hypothesis_tests}
\end{table}

\begin{table}
    \centering
    \begin{tabular}{lrr}
    \hline
         {\bfseries Method} & {\bfseries Mean Recall} & {\bfseries Dunn's test Rank} \\
         \hline
         \hline
         SZZ\_U & 0.99 & 1\\
         SZZ\_U+ & 0.99 & 1\\
         SZZ\_B+ & 0.97 & 2\\
         SZZ\_RA & 0.97 & 2\\
         SZZ\_RA+ & 0.97 & 2\\
         SZZ\_B & 0.96 & 2\\
         Proportion Moving Window & 0.81 & 3\\
         Proportion Increment & 0.76 & 3\\
         Proportion Cold Start & 0.76 & 3\\
         Simple & 0.36 & 4\\
         \hline
         \hline
    \end{tabular}
    \caption{Dunn's test results for the {\em recall} values of hypothesis $H_{20}$}.  
    \label{tab:rq3:recall_hypothesis_tests}
\end{table}

\begin{table}
    \centering
    \begin{tabular}{lrr}
    \hline
         {\bfseries Method} & {\bfseries Mean F1} & {\bfseries Dunn's test Rank}  \\
         \hline
         \hline
         Proportion Moving Window & 0.83 & 1\\
         Proportion Increment & 0.82 & 1\\
         Proportion Cold Start & 0.81 & 1\\
         SZZ\_B+ & 0.52 & 2\\
         SZZ\_B & 0.51 & 2\\
         SZZ\_RA & 0.51 & 2\\
         SZZ\_RA+ & 0.51 & 2\\
         SZZ\_U+ & 0.51 & 2\\
         SZZ\_U & 0.51 & 2\\
         Simple & 0.50 & 2\\
         \hline
         \hline
    \end{tabular}
    \caption{Dunn's test results for the $F1$ values of hypothesis $H_{20}$}.  
    \label{tab:rq3:f1_hypothesis_tests}
\end{table}

Our statistical results (i.e., Kruskal--Wallis and Dunn's tests) reveal that hypothesis $H_{20}$ can be rejected. Therefore, our proportion based methods for labeling defective classes significantly outperform the studied SZZ based methods in terms of the studied accuracy metrics (i.e., precision, recall, F1, kappa, and MCC). Tables~\ref{tab:rq3:precision_hypothesis_tests},~\ref{tab:rq3:recall_hypothesis_tests},~and~\ref{tab:rq3:f1_hypothesis_tests} show our results yielded by the Dunn's tests. Indeed, we observe that all the proportion based methods fall in the 1st rank of $F1$, whereas the SZZ based methods and the {\em Simple} method fall in the 2nd rank of $F1$, obtaining a statistically significant lower performance.

When comparing the proportion based methods, we note that the {\em Proportion Increment} method yields precision values that are statistically similar to the {\em Simple} method, which fell in the 1st rank. On the other hand, the {\em Proportion Cold Start} and {\em Proportion Moving Window} methods yield precision values that are statistically lower compared to the {\em Simple} method. Therefore, the {\em Proportion Increment} method likely produces better precision values than the {\em Proportion Cold Start} and {\em Proportion Moving Window} methods.

\subsection{RQ4: \rqfour}
\subsubsection{Design}
\label{sec:DesRQ4}

We propose the following hypothesis: 

{\em $\bullet H_{30}$: different methods lead to the same level of accuracy for feature selection.}

\textbf{Independent variables}

The independent variable is represented by the same methods used in RQ2 and RQ3. In RQ4, we use the labeled classes in RQ3 to select features.  

\textbf{Dependent variables}

Our dependent variable is the accuracy for selecting features \cite{Hall1998, kondo2019impact}. We compare which features are selected on the same dataset when created by our different studied methods.  The following metrics have been redefined for this RQ:
\begin{itemize}
    \item True Positive(TP): The feature is selected in the actual repository  and it is selected in the repository generated by a method.
    \item False Negative(FN): The feature is selected in the actual repository  and it is not selected in the repository generated by a method.
    \item True Negative(TN): The feature is not selected in the actual repository  and it is not selected in the repository generated by a method..
    \item False Positive(FP):  The feature is not selected in the actual repository  and it is selected in the repository generated by a method.
\end{itemize}

As features, to be selected, we used 17 well-defined product and project features that have been shown to be useful for defect prediction \cite{Falessi:2017:INS:3200492.3200504,DAmbros:2012:EDP:2318097.2318149}. Table \ref{table:Tablerq3a} details the set of features. 

\begin{table}[htb!]
	\begin{minipage}{.6\linewidth}
		\caption{Defect prediction features.}
		\centering
		\label{table:Tablerq3a}
		\begin{tabular}{c c}
			\begin{filecontents*}{Tables/Table4.csv}
Metric,Description
Size,Lines of code(LOC)
LOC Touched,Sum over revisions of LOC added + deleted
NR,Number of revisions
Nfix,Number of bug fixes
Nauth,Number of authors
LOC Added,Sum over revisions of LOC added
MAX LOC Added,Maximum over revisions of LOC added
AVG LOC Added,Average LOC added per revision
Churn,Sum over revisions of added - deleted LOC
Max Churn,Maximum churn over revisions
Average Churn,Average churn over revisions
Change Set Size,Number of files committed together
Max Change Set,Maximum change set size over revisions
Average Change Set,Average change set size over revisions
Age,Age of Release
Weighted Age,Age of Release weighted by LOC touched
\end{filecontents*}
\csvautotabular{Tables/Table4.csv}
		\end{tabular}
	\end{minipage}%
\end{table}

\textbf{Measurement Procedure}

\begin{figure}[H]
\begin{center}
   \includegraphics[width=1.0\columnwidth]{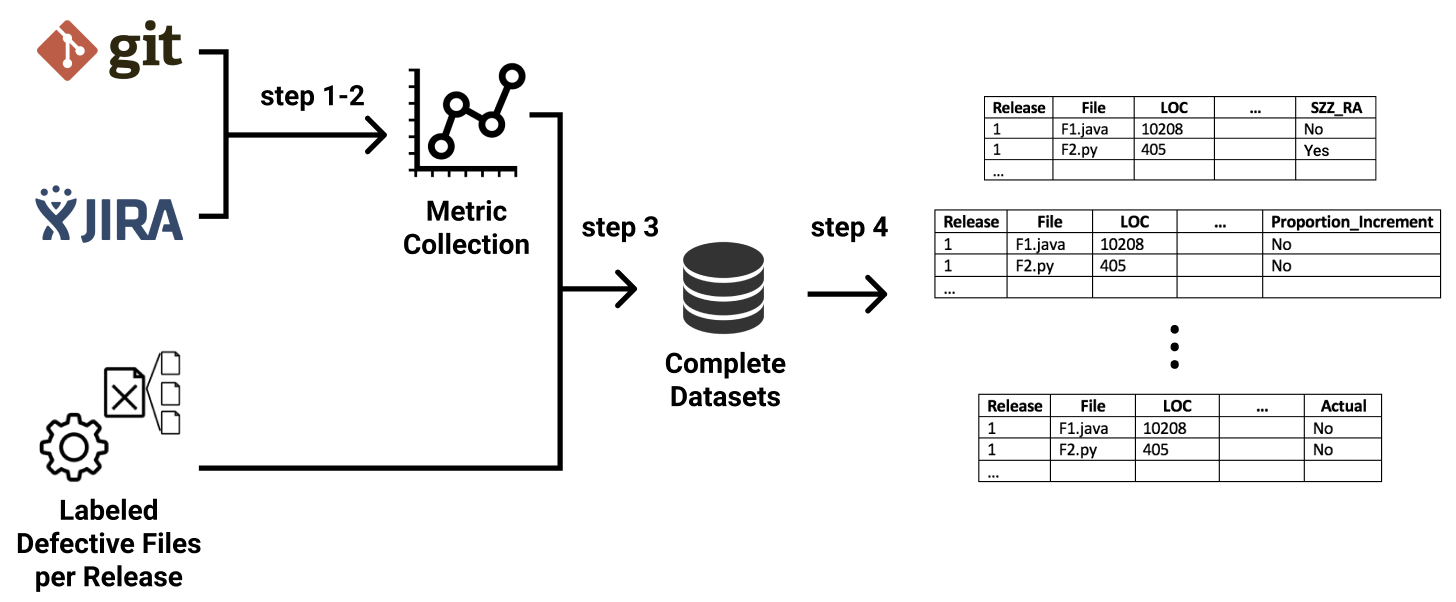}
  \vspace{-1mm}
    \caption{The process of creating the Complete datasets for each project and method.}
   \label{fig:rq4dataset}
   \vspace{-3mm} 
\end{center}
\end{figure}

For each project we compute the features in Table \ref{table:Tablerq3a} as shown in Figure \ref{fig:rq4dataset} and detailed in four steps.
\begin{enumerate}
    \item For each project, we begin by removing the last 50\% of versions due to the fact that classes snore as described by \citet{AalokSnoring2019}.
    \item For each project P, we compute the features as described in Table \ref{table:Tablerq3a} for each version-class pair. 
    \item For each of the methods M, we combined their produced AV datasets with the version-class pair's defectiveness (as computed in RQ3), which we labeled as P\_M\_Complete.
    \item For each version R within a project, we created a dataset including all version-class pairs with versions 1 to R labeled P\_M\_R\_Complete. This dataset uses the defectiveness computed by method M in RQ3.
    \end{enumerate}

\begin{figure}
\begin{center}
   \includegraphics[width=0.9\columnwidth]{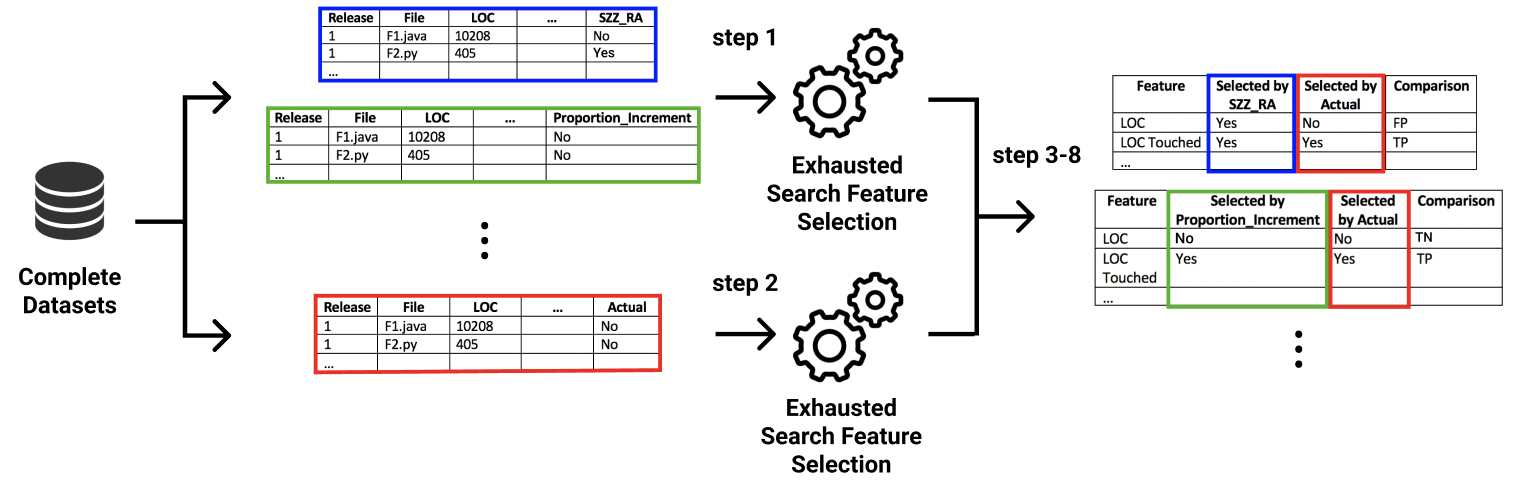}
  \vspace{-1mm}
    \caption{The process to measure the accuracy of methods in leading to accurate feature selection.}
   \label{fig:featureselection}
   \vspace{-3mm} 
\end{center}
\end{figure}

Afterwards, we analyze which features are selected. Figure \ref{fig:featureselection} reports an overview of the approach used in this RQ to measure the accuracy of methods for accurate feature selection. This approach consists of three steps: 

\begin{enumerate}
    \item For each dataset P, for each version R, we set the class defectiveness according to each method M, and we perform on P\_M\_R an Exhaustive Search Feature Selection\footnote{https://weka.sourceforge.io/doc.packages/attributeSelectionSearchMethods/} using Weka \cite{WEKAbook, kondo2019impact}. This search technique performs an exhaustive search through the space of features subsets starting from the empty set of features. If two subsets have the same  merit which are also the best merit encountered, then the technique favours the smaller subset. We used CfsSubsetEval\footnote{https://weka.sourceforge.io/doc.dev/weka/attributeSelection/CfsSubsetEval.html} for the evaluation function which evaluates the worth of a subset of features by considering the individual predictive ability of each feature along with the degree of redundancy between them. Subsets of features that are highly correlated with the class while having low inter-correlation are preferred \cite{Hall1998, kondo2019impact}.

    \item For each dataset P, for each version R, we set the class defectiveness according to the available actual/observed AVs and we perform, on P\_Actual\_R the Exhaustive Search Feature Selection using Weka and CfsSubsetEval as we did for the studied methods.
    \item For each P\_M\_R, we compare the features selected in P\_M\_R to the features selected in P\_Actual\_R.
\end{enumerate}

\textbf{Hypothesis testing.} To test hypothesis $H_{30}$, we use the same statistical machinery used in RQ2 and RQ3. We perform Kruskal-Wallis tests followed by Dunn tests.

\subsubsection{RQ4: Results}
\label{sec:ResRQ4}
\begin{figure}
  \includegraphics[width=0.6\columnwidth]{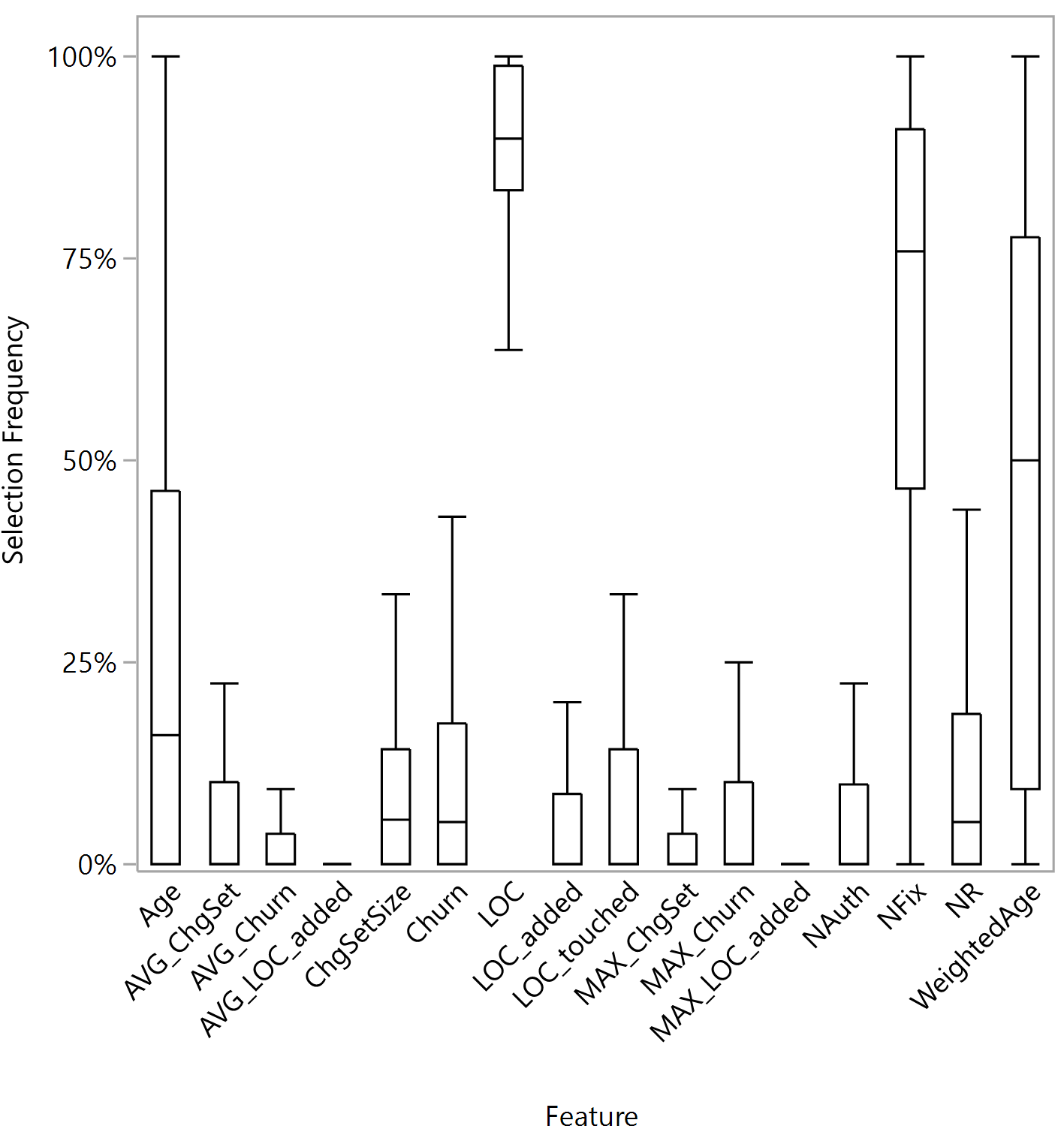}
  \vspace{-1mm}
    \caption{Distribution among datasets of selection frequency of each feature.}
   \label{fig:SelectionFrequency}
   \vspace{-3mm}
\end{figure}

Fig. \ref{fig:SelectionFrequency} reports the distribution among datasets of the actual selection frequency of each feature. Since the frequency of selection varies across features, then it is important to select the correct set of features.  

\begin{figure}
  \includegraphics[width=0.9\columnwidth]{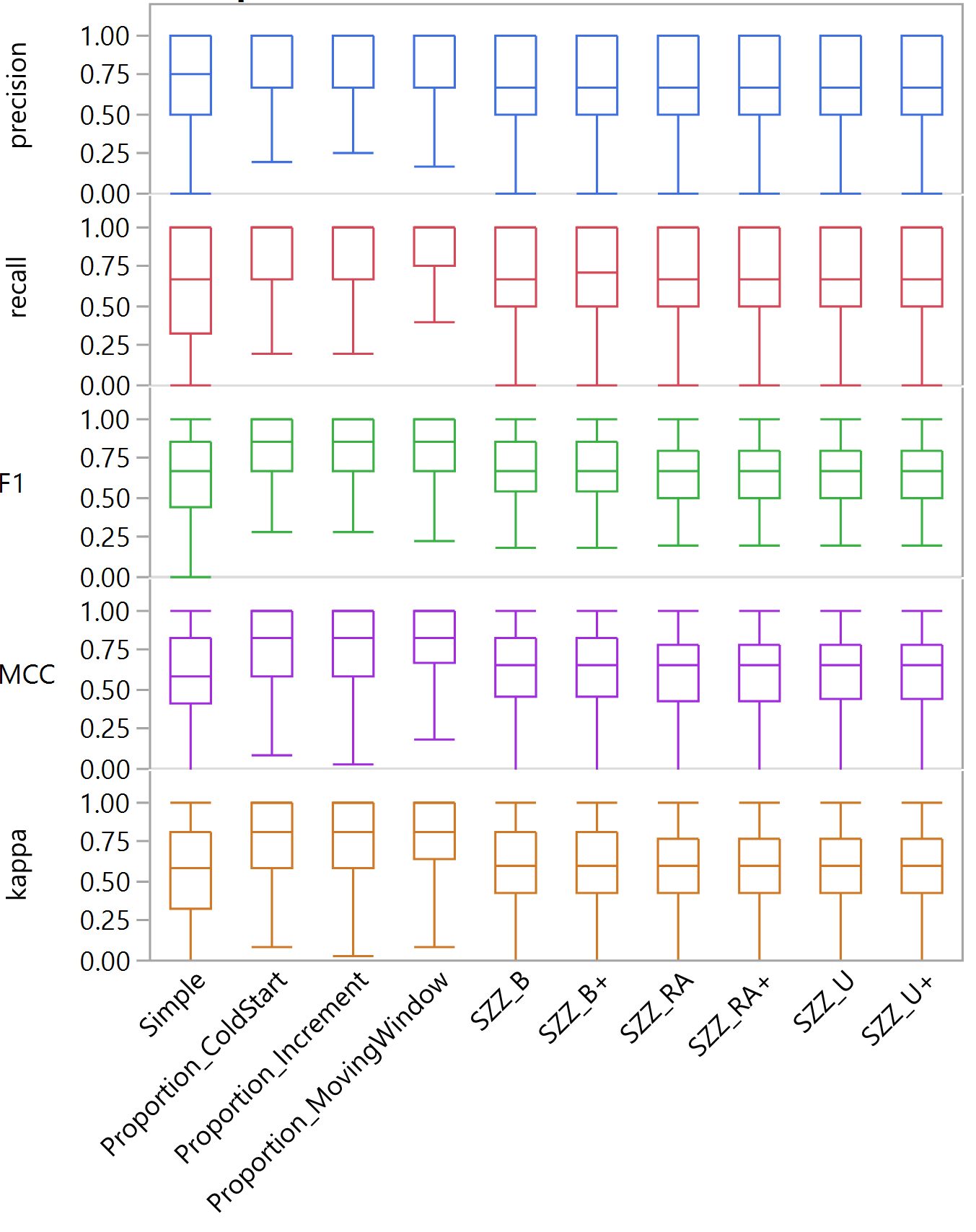}
  \vspace{-1mm}
    \caption{Distribution, across versions and 76 Apache projects, of Precision, Recall, F1, MCC, and Kappa, of different methods in feature selection.}
   \label{fig:SelectionAccuracy}
   \vspace{-3mm}
\end{figure}

Regarding the comparison of the features selected on a dataset produced by a method (i.e., where the AVs are those retrieved by a method) versus the features selected by using the actual/observed dataset (i.e., where the AV are those provided be developers), Figure \ref{fig:SelectionAccuracy} reports the distribution of a certain method to retrieve AVs (x-axis, across versions and 76 Apache projects) of Precision, Recall, F1, MCC, and Kappa for selecting features. According to Figure \ref{fig:SelectionAccuracy}, the proportion based methods have a higher accuracy (in all five metrics) compared to the studied SZZ methods. For example, according to Figure \ref{fig:SelectionAccuracy}, the proportion based methods are the only methods having a perfect median Precision and Recall.

Indeed, our Kruskall--Wallis and Dunn's tests reveal that hypothesis $H_{30}$ (i.e., different methods have the same accuracy when selecting features) can be rejected. Therefore, we can claim that \textbf{retrieving AVs based on the defects' life cycle can lead to an overall, and on average, more accurate feature selection than the studied SZZ methods.}

\begin{table}
    \centering
    \begin{tabular}{lrr}
    \hline
         {\bfseries Method} & {\bfseries Mean Precision} & {\bfseries Dunn's test Rank}  \\
         \hline
         \hline
         Proportion Moving Window & 0.84 & 1\\
         Proportion Cold Start & 0.83 & 1\\
         Proportion Increment & 0.82 & 1\\
         Simple & 0.74 & 2\\
         SZZ\_B+ & 0.70 & 3\\
         SZZ\_B & 0.70 & 3\\
         SZZ\_U+ & 0.69 & 3\\
         SZZ\_U & 0.69 & 3\\
         SZZ\_RA & 0.69 & 3\\
         SZZ\_RA+ & 0.68 & 3\\
         \hline
         \hline
    \end{tabular}
    \caption{Dunn's test results for the {\em precision} values of hypothesis $H_{30}$}.  
    \label{tab:rq4:precision_hypothesis_tests}
\end{table}

\begin{table}
    \centering
    \begin{tabular}{lrrl}
    \hline
	    {\bfseries Method} & {\bfseries Mean Recall} & {\bfseries Dunn's test Rank} & {\bfseries Comments} \\
         \hline
         \hline
	    Proportion Moving Window & 0.84 & 1 &\\
	    Proportion Cold Start & 0.82 & 1 &\\
	    Proportion Increment & 0.81 & 1.5 & Significantly lower than Proportion Moving Window\\
	    SZZ\_B+ & 0.71 & 2 &\\
	    SZZ\_B & 0.71 & 2 &\\
	    SZZ\_RA & 0.70 & 2 &\\
	    SZZ\_U & 0.70 & 2 &\\
	    SZZ\_RA+ & 0.70 & 2 &\\
	    SZZ\_U+ & 0.70 & 2 &\\
	    Simple & 0.61 & 3 &\\
         \hline
         \hline
    \end{tabular}
    \caption{Dunn's test results for the {\em recall} values of hypothesis $H_{30}$}   
    \label{tab:rq4:recall_hypothesis_tests}
\end{table}

\begin{table}
    \centering
    \begin{tabular}{lrrl}
    \hline
	    {\bfseries Method} & {\bfseries Mean F1} & {\bfseries Dunn's test Rank} & {\bfseries Comments}  \\
         \hline
         \hline
	    Proportion Moving Window & 0.83 & 1 &\\
	    Proportion Cold Start & 0.81 & 1 &\\
	    Proportion Increment & 0.80 & 1.5 & Significantly Lower than Proportion Moving Window\\
	    SZZ\_B+ & 0.67 & 2 &\\
	    SZZ\_B & 0.67 & 2 &\\
	    SZZ\_RA & 0.66 & 2 &\\
	    SZZ\_U & 0.66 & 2 &\\
	    SZZ\_U+ & 0.66 & 2 &\\
	    SZZ\_RA+ & 0.66 & 2 &\\
	    Simple & 0.64 & 2 &\\
         \hline
         \hline
    \end{tabular}
    \caption{Dunn's test results for the $F1$ values of hypothesis $H_{30}$}.  
    \label{tab:rq4:f1_hypothesis_tests}
\end{table}

Tables~\ref{tab:rq4:precision_hypothesis_tests},~\ref{tab:rq4:recall_hypothesis_tests},~and~\ref{tab:rq4:f1_hypothesis_tests} show the Dunn's test results. Indeed, we observe that, in terms of $F1$, the proportion based methods fall in better ranks than all the SZZ based methods and the Simple method. However, we also note that the {\em Proportion Increment} yields significantly lower Recall and F1 values than the {\em Proportion Moving Window method}.


\section{Discussion}\label{sec:discussion}
This section discuss our main results, the possible explanations for the results,  implications, and guidelines for practitioners and researchers.

\subsection{Main results and possible explanations}

The main implications of RQ1 is that most of the defects, of the 212 open-source projects from the Apache ecosystem, do not report AVs. However, according to Figure \ref{fig:RQ1v1}, the median project has most of the defects providing AVs. This means that in projects having a higher number of defects there is a higher proportion of missing AVs compared to projects having a small number of defects.

The main result of RQ2 and RQ3 is that \textbf{all proportion based methods have a higher precision and composite accuracy (F1, MCC, and Kappa) than all SZZ based methods}. One of the possible reasons for the high accuracy achieved by the proportion based methods is that P is substantially stable across projects (i.e., Proportion\_ColdStart) and more stable within the same project (i.e., Proportion\_Increment and Proportion\_MovingWindow).

\begin{figure}
  \includegraphics[width=0.6\columnwidth]{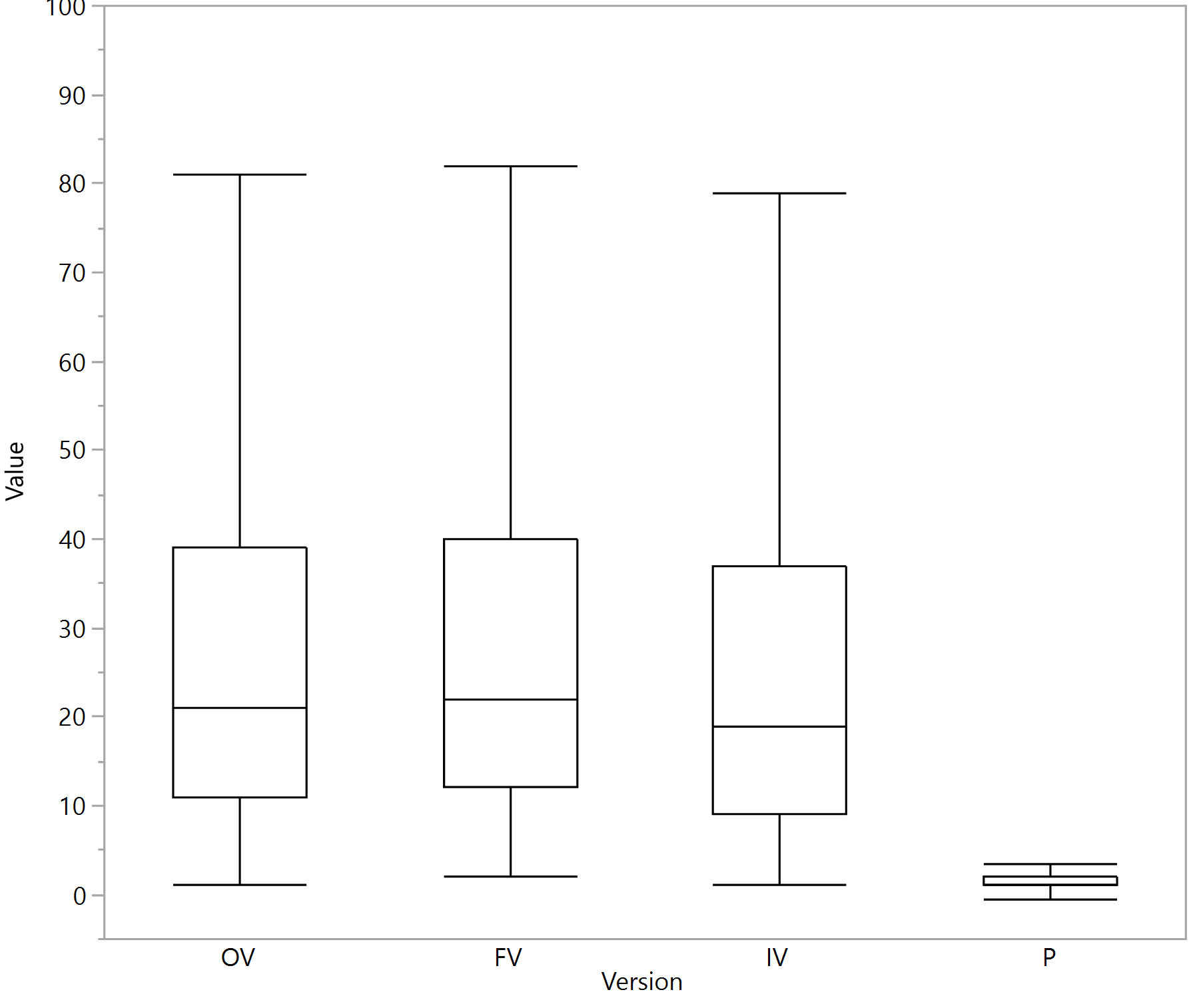}
  \vspace{-1mm}
    \caption{Distribution of values of IV, OV, FV, and P across defects of 76 Apache projects.}
   \label{fig:stability}
   \vspace{-3mm}
\end{figure}

Figure \ref{fig:stability} reports the distribution of values of IV, OV, FV, and P across defects of different projects. Table \ref{table:RQ2Stability} reports the variation, in terms of standard deviation, of IV, OV, FV, and P when it is computed across different projects. According to both Figure \ref{fig:stability} and Table \ref{table:RQ2Stability}, P is substantially stable across defects of different projects especially when compared to IV, OV and FV.

\begin{table} 
\caption{Variation, in terms of standard deviation (STDV), of IV, OV, FV, and P across defects of 76 Apache projects. }
\centering
\label{table:RQ2Stability}
\begin{tabular} { c c c }
\begin{filecontents*}{Tables/StabilityAcrossProjects.csv}
Version,STDV
IV,38.36
OV,40.17
FV,41.85
P,5.43
\end{filecontents*}
\csvautotabular{Tables/StabilityAcrossProjects.csv}
\end{tabular}
\end{table} 

\begin{figure}
  \includegraphics[width=0.6\columnwidth]{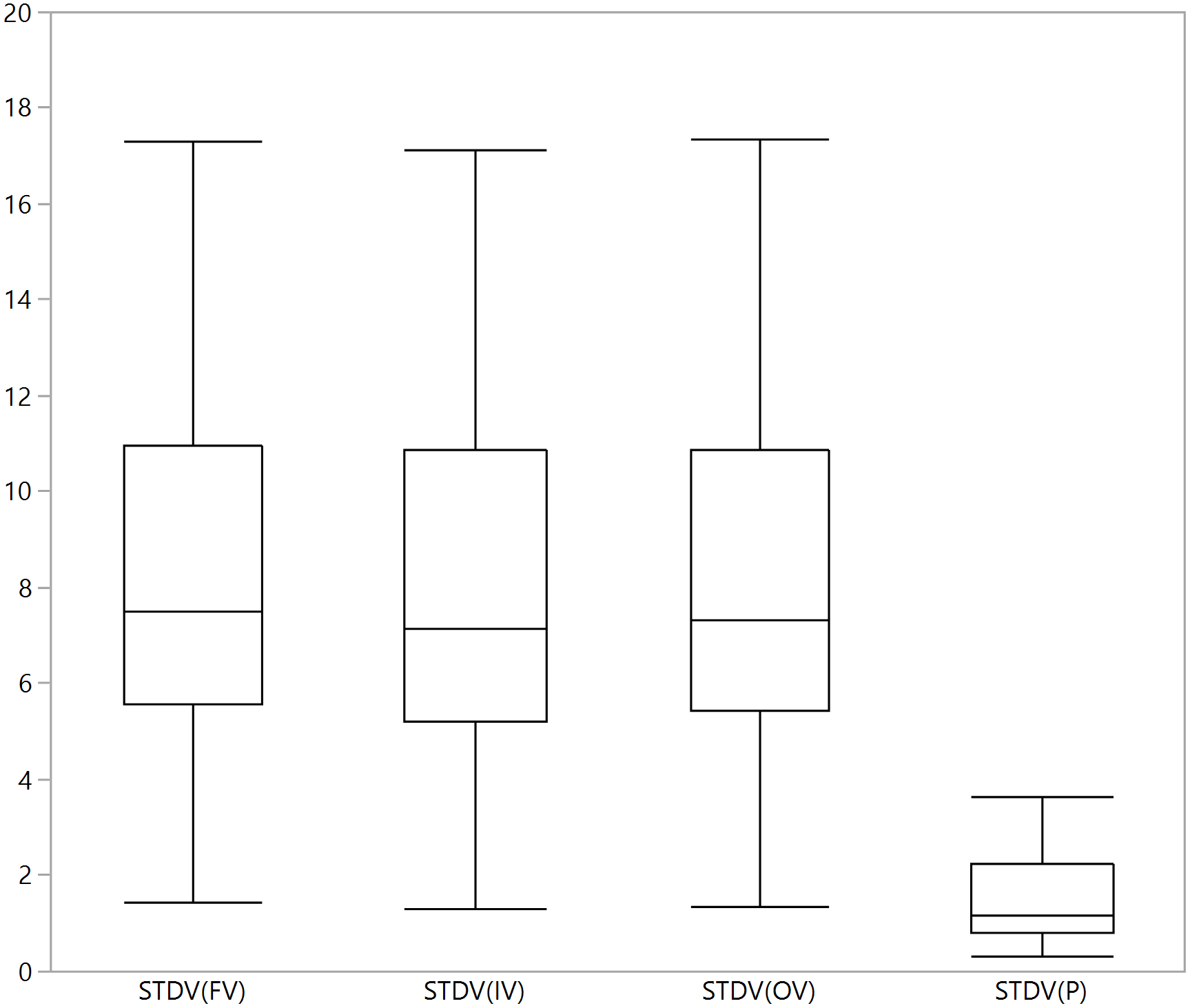}
  \vspace{-1mm}
    \caption{Distribution of standard deviation of IV, OV, FV, and P, as computed within the same project, across 76 Apache projects.}
   \label{fig:RQ2StabilitySTDV}
   \vspace{-3mm}
\end{figure}

An additional relevant results of RQ3 is that \textbf{Proportion\_MovingWindow method dominates all methods on all composite accuracy metrics (i.e., F1, MCC, and Kappa)}. This results is likely due to the fact that P is more stable within the same project than across projects. 
Figure \ref{fig:RQ2StabilitySTDV} reports the distribution of standard deviation of IV, OV, FV, and P, across 76 Apache projects. Standard deviation is one way of measuring the amount of variation of a set of values \cite{DBLP:journals/technometrics/Yeh05}. A low STDV indicates that the values tend to be close to the mean and hence spread out over a narrow range.
According to Figure \ref{fig:RQ2StabilitySTDV} the STDV is much higher across projects than within the same project. Specifically, the median STDV of P computed within the same project is less than 2 (Figure \ref{fig:RQ2StabilitySTDV}) whereas the one across projects is about 5 (Table \ref{table:RQ2Stability}). In conclusion, Figure \ref{fig:RQ1v1}, Table \ref{table:RQ2Stability} and Figure \ref{fig:RQ2StabilitySTDV} (STDV $<$2) show that \textbf{the proportion of number of versions between their discovery and their fix is more stable within the same project than across different projects.}

The main results of RQ4 is that \textbf{the proportion based methods have a higher accuracy (in all five metrics) than all  SZZ methods}. Moreover, by observing  Figure \ref{fig:SelectionAccuracy} we note that the accuracy of SZZ methods can reach even a negative value of kappa. This means that a random selection of features is more accurate than a selection based on a dataset produced by an SZZ based method.

RQ2, RQ3, and RQ4 share several results including that all proportion based methods have a higher precision and composite accuracy than all SZZ based methods. 
The major differences among RQ2 and RQ3 results is in that SZZ\_B+ has a higher composite accuracy than Simple and any other SZZ method. One possible reason is that, in RQ3, a class can be affected by multiple defects and therefore methods which miss defects can still perform accurately (see discussion in Section 3.3.2). By comparing Figure \ref{fig:RQ3result} to Figure \ref{fig:RQ2result} we observe that, all methods are more accurate for labeling classes (RQ3) than AVs (RQ2) on all accuracy metrics. Specifically, by comparing the median accuracy (across methods and datasets), we observe an increase in labeling classes over AVs of 13\% in Precision, 5\% in Recall, 16\% in F1,	27\% in MCC and	39\% in Kappa. It is interesting to note that the increase is higher in composite accuracy metrics than in atomic metrics. Again, we believe that one of the possible reasons is that, in RQ3, a class can be affected by multiple defects and, therefore, methods which miss defects can still perform accurately.

By comparing RQ4 to RQ2 and RQ3 we observe that there is less variation among accuracy of methods in RQ4 than in RQ2 or RQ3. In other words, the choice of the methods to retrieve AVs has less impact on feature selection (RQ4) than on class labeling (RQ3). However, in RQ2 and RQ3 the proportion based methods performed better than the other methods in four and three metrics, respectively, whereas in five metrics in RQ4. In other words, the superiority of the proportion based methods is clearer in feature selection (RQ4) than in class labeling (RQ3).
 Another major difference between RQ4 to RQ2 and RQ3 is that the distribution of accuracy is much wider in RQ4 than in RQ2 or RQ3. For instance, when the methods are inaccurate they are extremely less accurate in RQ4 than in RQ2 or RQ3. In other words, a medium amount of inaccuracy in class labeling leaded to a big amount of inaccuracy in feature selection. For example, the lowest score of Proportion\_MovingWindow in feature selection in Kappa, F1, Precision and Recall is less than 0.25 in RQ4 but higher than 0.6 in RQ3.
Similarly the lowest scores of SZZ based methods are even negative in case of Kappa for feature selection (RQ4) and higher than 0.2 in RQ3.

As different methods suggest different affected classes, then it could be possible to improve the accuracy of single methods in labeling defective classes by combining them through machine learning models. We tried several approaches and our preliminary results were negative as they showed that feature selection---performed by using a machine learning model like random forest to merge the methods---identified Proportion\_MovingWindow as the only important method. Thus, in the future we plan to experiment with combining more methods, on different datasets, by using different feature selection techniques and machine learning models.
Finally, someone could think that SZZ can be used more widely than Proportion as SZZ is decoupled from the development process used in a specific context and from possible changes in that process. However, the fact that Proportion\_ColdStart outperforms SZZ suggests the opposite.

One of the possible reasons why SZZ methods have a higher Recall than proportion based methods is that SZZ based methods inflate the number of AVs as they produce a substantially high number of defect-inducing changes when compared to the ones produced by the proportion based methods. The results regarding F1 clearly show the cost of this inflation.

It could be that other window lengths outperform our current 1\% window length. Moreover, it could even be that the best window length changes across projects or even within the same project. Thus, in future work, we plan to investigate how to dynamically define the moving window length across and within projects.

\subsection{Implications}

The main implications of RQ1 is that relying on the available AVs means neglecting most of the defects. Therefore, effort should be invested in retrieving AVs (hence the importance of our work).

The main implications of RQ2 and RQ3 to practitioners is that SZZ should be used over proportional methods only in cases where higher Recall values are preferred over Precision, F1, MCC, and Kappa values. As for researchers, the results of RQ2 highlights exciting possibilities for future work in the area of defect introduction. For example, the framework proposed by \citet{CostaMSKCH17} to evaluate SZZ implementations can be enhanced with the AVs retrieved by our proportion-based methods.  
The main implications of RQ4 results to practitioners and researchers is to prefer using the proportion based methods over SZZ based methods when performing feature selection. Specifically, practitioners can use our proportion based methods and, more generally, our results in four different ways: 
\begin{itemize}
\item Supporting decisions: the most obvious way to use proportion methods by practitioners is the one envisioned in RQ4; i.e., the practitioners can mine the dataset developed via proportion methods to understand what correlates with class defectiveness and make decisions according to this \cite{Zimmermann:2007:PDE:1268984.1269057, Hall1998,DBLP:journals/sigsoft/SliwerskiZZ05, DBLP:conf/msr/EyolfsonTL11, DBLP:conf/msr/AsaduzzamanBRS12, DBLP:conf/icse/RahmanD11, DBLP:conf/msr/WeissPZZ07, DBLP:conf/csmr/BernardiCLPD12,DBLP:conf/msr/RahmanBD10,DBLP:conf/icse/KimZWZ07}. For instance, in a previous paper \cite{DBLP:journals/software/FalessiSM14}, the third author reported on a custom Web-based tool called the Measurement Reporting Tool used in a software company called Keymind  to support decisions about software releases based on the predicted number of defects that a version would have. Possible decisions include the level of compliance of the code to the architecture and the number of requirements to implement in that version. If the datasets to mine would have been developed via SZZ, then this would lead us to the use of the wrong features, incurring wrong decisions regarding versions, thus incurring more defects, and hence more failures. This would ultimately lead to an economic loss of the company. 
\item Supporting testing: an additional way to use proportion methods is to mine the dataset developed via proportion methods for predicting which class is prone to be defective. During testing, the developers can focus on classes that are prone to be defective and hence avoiding more defects given a limited amount of effort resources. However, if the datasets to mine would have been developed via SZZ, then this would lead us to the use of the wrong features. These wrong features would in turn lead to inaccurate prediction of defect prone classes, which would, in turn, as explained before, ultimately lead to an economic loss of the company.
\item Supporting JIRA: a further way to use our proportion based methods is to be installed in an issue tracking system like JIRA and, when the user would have to provide the AV information, the tool could suggest the potential AVs as retrieved by the Proportion\_MovingWindow. The tool could also provide a warning message asking the user to confirm the submitted AV if it is very different from the suggested AV. 

\item Debugging: finally, practitioners can use the AVs retrieved by Proportion\_MovingWindow for debugging activities. Given that a new defect is reported and awaiting for a fix, the related retrieved IV could be used to understand the context in which the defect has been introduced, e.g., the developers could think of the features introduced in that IV as the potential source of the defect. Helping developers to narrow down to the version when a defect was introduced would considerably help to find the mistake to be fixed.
\end{itemize}

The overall main implication to researchers is twofold: 
\begin{enumerate}
\item the need to revisit the studies that have used SZZ methods as the oracle upon which the studies have based their conclusions \cite{DBLP:journals/tse/Kim/2008, DBLP:journals/tse/Kamei/2012, DBLP:conf/icsm/KameiMMMAH10, DBLP:conf/msr/Fukushima/2014}. 
\item the need to provide more accurate methods as no method to label affected versions or defective classes is perfect. The Kappa in Figure 5 and Figure 7 is far from 1.0 and therefore future studies are needed. 
\end{enumerate}

Finally, we would like to propose some guidelines for researchers and practitioners. Researchers should refrain from using SZZ for creating oracles. Researchers should create the oracles by using the available AV and, when missing, they should use the Proportion\_MovingWindow method. Since no proportion method is perfect, and likely no method will ever be perfect, researchers should prefer mining projects with a high proportion of available and consistent AV. Practitioners should carefully report the AV in all defect reports, they could be aided by the proportion based method in case they believe providing AV is complex.

\section{Threats to Validity}\label{sec:threats}
In this section, we report the threats to validity of our study. The section is organized by threat type, i.e., Conclusion, Internal, Construct, and External.

\subsection{Conclusion}
Conclusion validity concerns issues that affect the ability to draw accurate conclusions regarding the observed relationships between the independent and dependent variables \cite{Wohlin:2012}.

We tested all hypotheses with non-parametric tests (e.g., Kruskal--Wallis) which are prone to type-2 error, i.e,. not rejecting a false hypothesis. We have been able to reject the hypotheses in most of the cases; therefore, the likelihood of a type-2 error is low. Moreover, the alternative would have been using parametric tests (e.g., ANOVA) which are prone to type-1 error, i.e., rejecting a true hypothesis, which in our context is less desirable than type-2 error. Also, we acknowledge that our proposed methods (i.e., independent variables) do not represent an exhaustive list of methods that could have been implemented (for example, one could use machine learning to optimize the proportions used in the ColdStart method). However, our proposed methods are a simple and effective baseline to start with (as shown by our obtained results).

\subsection{Internal}\label{sec:threats_internal}
Internal validity is concerned with the influences that can affect the independent variables with respect to causality \cite{Wohlin:2012}.
A threat to {\em internal validity} is the lack of ground truth for class defectiveness, which could have been underestimated in our measurements. In other words, the AVs provided by developers might be inaccurate due to human error. Nevertheless, we would argue that this is a common threat in most of empirical research in the area of software engineering~\cite{kamei2016defect}.

\subsection{Construct}
Construct validity is concerned with the degree to which our measurements indeed reflect what we claim to measure~\cite{Wohlin:2012}.  

In our study, we compare our proposed proportion methods with the SZZ based methods. We are aware that the output of SZZ are defect-introducing changes and not affected versions. For example, although SZZ may output three distinct defect-introducing changes (which we may interpret as three distinct affected versions), we do not investigate the dependency between these defect-introducing changes. For instance, a defect may only be present when all the three defect-introducing changes are present. Therefore, a version that contains only one of the defect-introducing changes may not be, in actuality, an affected version. Nevertheless, our assumptions are aligned with prior work, which has considered every potential defect-introducing change as indeed defect-introducing~\cite{CostaMSKCH17} and, therefore, can be interpreted as incurring an affected version.

Moreover, we use Precision, Recall, F1-Score, Matthews Correlation Coefficient, and Cohen's Kappa to measure the accuracy for labeling defectiveness in RQ2 and RQ3. Although we do not use the {\em Area Under the Curve} (AUC) metric, which is a threshold-free metric~\cite{DBLP:journals/tse/Tantithamthavorn/2016}, our methods do not output probabilities. Therefore, our evaluations are not impacted by threshold choices.
%

\subsection{External}
External validity is concerned with the extent to which the research elements (subjects, artifacts, etc.) are representative of actual elements \cite{Wohlin:2012}.

This study used a large set of datasets and hence could be deemed of high generalization compared to similar studies. Of course, our results cannot be generalized by projects that would significantly differ from the settings used in this present study.

Finally, in order to promote reproducible research, all datasets, results and scripts for this paper are available in our replication package\footnote{\url{https://gitlab.com/Bvandehei/affectedversions}}.

\section{Conclusion}\label{sec:conclusion}

In this paper, we first measured the AV availability and consistency in open-source projects, and then evaluated a new method for retrieving AVs, i.e., the origin of a defect, which is based on the idea that defects have a stable life cycle in terms of proportion of number of versions required to discover and to fix the defect.
Our results regarding 212 open-source projects from the Apache ecosystem, featuring a total of about 125,000 defects, show that the AVs cannot be used in the majority (51\%) of defect reports. Therefore, it is important to develop automated methods to retrieve AVs. Results related to 76 open-source projects from the Apache ecosystem, featuring a total of about 6,250,000 classes that are are affected by 60,000 defects and spread over 4,000 versions and 760,000 commits, show that our proposed methods are, on average, more accurate  when compared to previously proposed and state-of-art SZZ based methods, for retrieving AVs. Our results suggest that our proposed methods are also better than SZZ based methods for labeling classes as defective and for developing defects repositories to perform feature selection. In conclusion, our proposed methods are a valid automated alternative to SZZ for estimating the origin of a defect and hence for building defects repository for defect prediction endeavours.

Future studies include:
\begin{itemize}    
    \item \textbf{Analyzing other defect-introducing commits in SZZ methods}. In our research, we selected the earliest possible defect-introducing commit returned by SZZ to be the IV for a defect. Future work will focus on how selecting later defect-introducing commits affects the accuracy in labeling classes in versions as defective or not. 
    \item \textbf{Analyzing the role of reporting affect versions to developers}. In our study, we only analyzed whether AVs were available and consistent. Future work will focus on why and how developers report AVs; how do developers determine AVs? Do developers find reporting AVs important?
    \item \textbf{Replication in context of JIT}. Just In Time (JIT) prediction models, where the predicted variable is the defectiveness of a commit, have become sufficiently robust that they are now incorporated into the development cycle of some companies\cite{McIntoshK18}. Therefore, it is important to investigate the accuracy of proportion based methods in the context of JIT models.
    \item \textbf{Finer combination of proportion based and SZZ based methods}. In this work, we have combined SZZ and proportion based method by simply tagging a version as defective if it came after the defect report creation and not tagged by SZZ. More finer combination are possible including the use of ML; i.e., the dataset to evaluate and use ML models can be created by ML models.
    \item \textbf{Use a finer P}. In this work, we simply used the proportion of versions to find and to fix a defect to determine P, which is then used to label AVs and classes. However, there exists room for improvement in calculating P. For example, P can be improved using Linear Regression. In addition to the version information, the number of days can also be used.
\end{itemize}

\bibliographystyle{ACM-Reference-Format}
\bibliography{bibliography}

\end{document}